\documentclass[11pt]{article}

\pdfoutput=1

\usepackage{amssymb}
\usepackage{amsmath}
\usepackage{mathrsfs} 
\usepackage[FIGTOPCAP]{subfigure}
\usepackage{graphicx}

\usepackage[hypertex]{hyperref}

\newcommand{ \centeron }[2]{{\setbox0=\hbox{#1}\setbox1=\hbox{#2}\ifdim
                             \wd1>\wd0\kern.5\wd1\kern-.5\wd0\fi \copy0
                             \kern-.5\wd0\kern-.5\wd1\copy1\ifdim\wd0>\wd1
                             \kern.5\wd0\kern-.5\wd1\fi}}
\newcommand{ \ltap }{\>\centeron{\raise.35ex\hbox{$<$}}
                     {\lower.65ex\hbox{$\sim$}}\>}
\newcommand{ \gtap }{\>\centeron{\raise.35ex\hbox{$>$}}
                     {\lower.65ex\hbox{$\sim$}}\>}

\newcommand{\lsim}{\lesssim}

\newcommand{\ra}{\rightarrow}

\newcommand{\neut}{N}
\newcommand{\charg}{C}
\newcommand{\hino}{\tilde{H}}
\newcommand{\bino}{\tilde{B}}

\begin{document}

\begin{titlepage}

\vspace*{1.5cm}

\centerline{\LARGE\bf $\mu$ to $e$ in $R$-symmetric Supersymmetry}

\vspace*{0.5cm}

\begin{center}
Ricky Fok and Graham D. Kribs \\
\vspace*{0.5cm}
{\it Department of Physics, University of Oregon, Eugene, OR, 97403} \\
\hspace*{0.5cm}
\end{center}

\vspace*{0.4cm}

\begin{abstract}

We demonstrate that $\mu \leftrightarrow e$ slepton mixing 
is significantly more restricted than previously thought within the 
already remarkably flavor-safe $R$-symmetric supersymmetric standard model.
We calculate bounds from $\mu \ra e \gamma$, $\mu \ra 3e$ and, 
most importantly, $\mu \ra e$ conversion.  
The process $\mu \ra e$ conversion is significantly more restrictive
in $R$-symmetric models since this process can occur through 
operators that do not require a chirality-flip.  
We delineate the allowed parameter space, demonstrating that 
maximal mixing is rarely possible with weak scale superpartners, 
while $\mathcal{O}(0.1)$ mixing is permitted within most of the space. 
The best approach to find or rule out $\mu \leftrightarrow e$ mixing in
$R$-symmetric supersymmetric models is a multi-pronged attack looking 
at both $\mu \ra e$ conversion as well as $\mu \ra e\gamma$.
The redundancy eliminates much of the parameter space where 
one process, but not both processes, contain amplitudes that
accidentally destructively interfere.  We briefly discuss 
implications for searches of slepton flavor violation at the LHC.

\end{abstract}

\end{titlepage}

\newpage
\setcounter{page}{2}

\section{Introduction}
\label{intro-sec}

Lepton flavor violation (LFV) is predicted to occur at an unobservably 
small rate in the Standard Model (SM).  In low energy supersymmetric
theories, new sources of lepton flavor violation are generic 
in the soft breaking sector.  The experimental non-observation of 
$\mu \rightarrow e$ processes is particularly restrictive, 
given the impressive 
bounds on $\mu \rightarrow e\gamma$ from MEGA \cite{Ahmed:2001eh}
and MEG \cite{Adam:2009ci}; on $\mu \rightarrow e$ conversion
from SINDRUM~II \cite{Bertl:2006up}, and to a lesser extent from 
$\mu \rightarrow 3e$ from SINDRUM \cite{Bellgardt:1987du}.
Further progress is expected from the varied experiments
that are ongoing as well as planned future experiments such as 
Mu2e \cite{Kutschke:2009zz} and other proposals utilizing
Project X at Fermilab \cite{projectx}.

In the minimal supersymmetric standard model (MSSM), 
$\mu \leftrightarrow e$ mixing is severely constrained by these bounds
(e.g. \cite{Hisano:1995cp,Hisano:1998fj,Masina:2002mv,Paradisi:2005fk,Ciuchini:2007ha}).
The size of the mixing can be characterized by the quantity 
$\delta^\ell_{XY} \equiv \delta m^2_{XY}/m^2$ where 
$\delta m^2_{XY}$ is the off-diagonal $(12)$-entry appearing in the
sfermion mass matrix connecting the $X$-handed slepton to the
$Y$-handed slepton, and $m^2$ is the average slepton mass.  
Ref.~\cite{Ciuchini:2007ha} found $\delta^\ell_{LR} \lsim 3 \times 10^{-5}$,
while $\delta^\ell_{LL} \lsim 6 \times 10^{-4}$ over a scan
of the mSUGRA parameter space.  Similarly strong bounds on 
$\delta^\ell_{RR}$ can also be found, though cancellations between 
diagrams in the amplitude can in some cases allow for much larger mixing
\cite{Masina:2002mv,Paradisi:2005fk,Ciuchini:2007ha}.

Recently, a new approach to weak scale supersymmetry that
incorporates an extended $R$-symmetry \cite{Kribs:2007ac}, 
suggests large flavor violation in the supersymmetry breaking 
parameters may be present \emph{without} exceeding the 
flavor-violating bounds.
This is possible for several reasons:
$R$-symmetric supersymmetry has no flavor-violating LR mixing,
solving the worst of the problem trivially.  
$R$-symmetric supersymmetry has Dirac gauginos, and 
no Majorana masses, removing all dimension-5 flavor-violating operators.
Finally $R$-symmetric supersymmetry also has no flavor-conserving 
LR mixing, and so there are no ``large $\tan\beta$ enhanced'' effects.
These benefits were found to virtually eliminate constraints
on the slepton flavor mixing \cite{Kribs:2007ac}. 

In this paper we reconsider the constraints on slepton mixing,
specifically, $\mu \leftrightarrow e$ mixing.  Unlike the MSSM,
the most important constraint is not necessarily $\mu \ra e\gamma$.
This is easily seen by inspection of the $R$-symmetric flavor-violating
operators:  $\mu \ra e\gamma$ requires a chirality-flip via a muon Yukawa
coupling, whereas $\mu \ra e$ conversion has no such requirement.
We find that $\mu \ra e$ conversion rules out maximal mixing 
throughout the right-handed slepton mixing parameter space for sub-TeV 
superpartner masses.  This is complementary to $\mu \ra e\gamma$,
where we find cancellations between the bino and Higgsino diagrams,
analogous to what was found before in the MSSM 
\cite{Masina:2002mv,Paradisi:2005fk,Ciuchini:2007ha}.
For left-handed slepton mixing, we find possible cancellations in the
amplitudes for $\mu \ra e$ conversion, and instead $\mu \ra e\gamma$
provides generally the strongest constraint.  We also calculated
$\mu \ra 3e$ and find it provides the weakest constraint on both 
left-handed and right-handed slepton mixing throughout the 
parameter space we consider.

This paper is organized in as follows:  We review the 
relevant characteristics of a model with an extended $R$-symmetry, 
and the super GIM mechanism in Sec.~\ref{simp-sec}.
In Sec.~\ref{exp-sec}, we begin the discussion of 
experimental constrains on the parameters from
$\mu \ra e \gamma$, in Sec.~\ref{mueg-subsec}, 
$\mu \ra e$ conversion in Sec.~\ref{muec-subsec}, and finally, 
$\mu \ra 3e$ in Sec.~\ref{mu3e-subsec}. 
In Sec.~\ref{lhc-sec} we briefly discuss implications for 
slepton flavor violation to be observed at LHC\@. 
Finally, in Sec.~\ref{discussion-sec} we conclude with a discussion
of our results.

\section{A Simplified R-symmetric Model}
\label{simp-sec}

We are interested in analyzing LFV in the minimal $R$-symmetric standard 
model (MRSSM).  The gaugino structure of the MRSSM has been studied
in detail in Ref.~\cite{Kribs:2008hq}, where the mixings and couplings
of the four Dirac neutralinos and four Dirac charginos are given.
Weak scale supersymmetry with Dirac gauginos is a possibility that was 
contemplated some time ago 
\cite{Fayet:1978qc,Polchinski:1982an,Hall:1990hq} 
and more recently
\cite{Fox:2002bu,Nelson:2002ca,Chacko:2004mi,Carpenter:2005tz,Antoniadis:2006uj,Hisano:2006mv,Hsieh:2007wq,Kribs:2007ac,Kribs:2008hq,Choi:2008pi,Amigo:2008rc,Plehn:2008ae,Harnik:2008uu,Benakli:2008pg,Luo:2008zr,Kribs:2009zy,Blechman:2009if,Kumar:2009sf}.
A fully general analysis of LFV in the MRSSM would be a substantial 
undertaking.  Fortunately, there are several simplifications we can
employ to gain a fairly general understanding of the allowed parameter
space of LFV in the MRSSM\@.  One important restriction is that the
Dirac wino cannot be light in the MRSSM, due to the structure of the 
wino supersoft operator \cite{Fox:2002bu}.  Essentially there is an 
unavoidable contribution to the vev of the $SU(2)_L$-triplet scalar that 
causes a contribution to the $\rho$ parameter that is too large unless the
wino is above about a TeV\@.  Secondly, since there is no coupling between
up-type Higgs and leptons, the contribution from the up-type Higgsino
eigenstates is suppressed by the small mixing between bino or 
$\hino_d$ and $\hino_u$, and so can be ignored. 

Itemizing the simplifications, we take:

\begin{itemize}

\item[1.] The wino mass, $M_2$, is taken to be sufficiently
large so as to give negligible contribution to flavor violating
interactions.  This simplification means that the $\rho$-parameter
is automatically safe throughout the parameter space we consider.

\item[2.] The up-type Higgsino mass $\mu_u$, is also taken
to be large for convenience.  Since the up-type Higgsinos play 
no role whatsoever in charged lepton flavor-violation (given also point 1), 
this is done simply to keep the gaugino sector 
to a $2\times2$ structure and thus easily understood. 
(We will, however, consider effects of a light up-type Higgsino 
on flavor-violating signals at LHC in Sec.~\ref{lhc-sec}.)

\item[3.] We consider left-handed and right-handed slepton mixing
separately.  This is standard practice when considering flavor-violation
in the MSSM (e.g., \cite{Ciuchini:2007ha}).  We will see that
there are qualitative differences between the allowed parameter
space of left-handed and right-handed slepton mixing.

\item[4.] We assume the slepton mixing is purely in the $2 \times 2$
flavor space of $e,\mu$.  Enlarging this mixing to the full
$3 \times 3$ mixing does not qualitatively change any of our results, 
and instead simply dilutes the effect of the mixing, while adding
more mixing angles and thus more parameters to the model. 
Since the focus of this paper is to explore $\mu \leftrightarrow e$
mixing, no further discussion of the $3 \times 3$ case will be given.

\item[5.] For our numerical results, we take 
$m_{\tilde{l}_2} = 1.5 m_{\tilde{l}_1}$.  This seems a far 
more drastic assumption than it actually is.  Our motivation is to 
consider slepton flavor violation when there is essentially
\emph{no degeneracy} among the sleptons, and so we took the
slepton mass ratio to be ``order one'' but not near one.
Taking the ratio much larger than one does not appreciably
increase the flavor violation, while taking it smaller causes
the super-GIM mechanism to suppress the flavor-violating signal.
Our compromise is the above number.  

\end{itemize}

In Appendix~\ref{app-sec}, we provide more details on the 
gaugino structure and flavor-violating interactions as directly
relevant to this paper.  With the above assumptions, there
is only one light Dirac chargino (which is $\hino_d^\pm$-like)
and two light Dirac neutralinos (which are mixtures of 
$\hino_d^0$ and $\bino$).

A few more comments on the slepton mass eigenstate hierarchy
are in order.  MSSM analyses of slepton flavor violation have,
by necessity of LFV constraints, concentrated on the case where
the mass difference between the different states is small, 
$\Delta^2 \equiv m_{\tilde{l}_2}^2 - m_{\tilde{l}_1}^2 
\ll m_{\tilde{l}_{1,2}}^2$.  In this limit, it is straightforward
to show that the contribution to LFV can be expanded in
powers of $\Delta^2$, taking the form 
\begin{eqnarray}
\sin 2 \theta_l \left( \frac{\Delta^2}{M_{\rm SUSY}^2} + \ldots \right) \; ,
\end{eqnarray}
where $M_{\rm SUSY}$ is typically the largest mass sparticle in the diagram
that dominates the process. There is no $\Delta$-independent contribution 
within the parentheses due to the super GIM mechanism (see the next section).  
Since $\sin 2\theta_l = 2 m_{e\mu}^2 / \Delta^2$,
one factor of $\Delta^2$ cancels, giving proportionality
to the $\delta$ parameter mentioned in the introduction
and used in many other papers on LFV in the MSSM 
(at least up to a possible further suppression of 
$|m_{\tilde{l}_1} m_{\tilde{l}_2}|/m_{N,C}^2$ if 
$m_{N,C} \gg m_{\tilde{l}_{1,2}}$).  

In this paper, $\Delta^2$ is not small, and so using the 
``$\delta$ parameter'' is simply not appropriate.  Instead, it is
easy to see that in the opposite limit, $\Delta^2 \gg m_{\tilde{l}_1}^2$, 
the contributions to LFV are proportional to simply
$\sin 2\theta_l/m_{\tilde{l}_1}^2$.
Hence, the relevant parameters we show in most of our 
numerical results are bounds on $\sin 2\theta_l$ as a function
of the slepton, gaugino, and Higgsino masses.
Reducing the splitting can be roughly approximated by relaxing 
the constraint on $\sin 2\theta_l$ by ratios of 
$\Delta_{\rm old}^2/\Delta_{\rm new}^2$.

\subsection{The super GIM mechanism}
\label{supergim-sec}

The ``super-GIM mechanism'' -- the GIM mechanism applied to flavor 
in the superpartner sector -- is important in understanding 
the phenomena of flavor violation.  As is well known, the super-GIM
mechanism arises as a consequence of the unitarity of the 
slepton mixing matrices that diagonalize the mass matrix; 
$U^{\dagger}_{ik}U_{kj}=\delta_{ij}$, where the sum 
over repeated indices is performed. This combination of mixing matrix 
elements always appears as a prefactor in the calculation of 
amplitudes of flavor violating processes. Specifically for our case 
of slepton flavor violation, we have $U^{\dagger}_{ek}U_{k\mu} = 0$, 
corresponding to an incoming muon, and an outgoing electron, 
with internal sleptons labeled by $k$.  The sum over $k$ corresponds 
to summing over all mass eigenstate sleptons $\tilde{l}_k$ in the loop. 
There are two immediate consequences of the super-GIM mechanism.

First, terms that do not depend on the slepton masses 
do not contribute.  Let $f(m_k)$ be some function that depends 
on the mass of the sleptons and $\alpha$ be some quantity that 
does not depend on $m_k$, then
\begin{equation}
\sum_k U^{\dagger}_{ek}U_{k\mu}[\alpha + f(m_k)] 
    = \sum_k U^{\dagger}_{ek}U_{k\mu} f(m_k).
\label{GIM}
\end{equation}
The form of Eq.~(\ref{GIM}) appears when a logarithmic divergent 
loop integral is dimensionally regularized, and one finds the 
$1/\epsilon$ term appearing as a constant term $\alpha$ in the 
above equation. This leads to an important result: the would-be 
logarithmic UV divergence in flavor-conserving processes 
is, in fact, UV finite in flavor-violating processes. 
In this paper, unless otherwise stated, we will omit the terms 
in our expressions that are canceled by the super-GIM mechanism.

The other well known consequence is that, when all the sleptons are 
degenerate, there is no flavor violation.  This can be seen again 
in Eq.~(\ref{GIM}) with $m_k = m$, the sum over all slepton flavors 
in a flavor-violating process vanishes.

\section{Experimental constraints}
\label{exp-sec}

There are three $\mu \to e$ conversion processes with experimental
bounds:  $\mu \ra e \gamma$, $\mu \ra e$ conversion, and 
$\mu \ra 3e$.  In this section we present our calculations of 
the rates of these processes and present results in terms of
a series of contour plots showing the allowed parameter space.

The rate for $\mu \ra e \gamma$ was estimated in Ref.~\cite{Kribs:2007ac}
in the slepton flavor-violating mass-insertion approximation with 
a pure bino and wino and a specific gaugino hierarchy.  
In this paper we have neglected the wino, due to the $\rho$ parameter
constraint, and instead included the down-type Higgsino $\hino_d^0$.
Since we have considered large mixing angles, up to and including
maximal mixing, we have diagonalized the slepton masses explicitly 
and done our loop calculations involving the slepton mass eigenstates.

As stated in our simplifications, we have not included contributions
from the wino or up-type Higgsino.  We focus on the case where 
the sleptons and the lighter neutralinos are in the sub-TeV range 
where wino contributions can be reasonably ignored. 
The up-type Higgsino does not couple to leptons, and we take
the light quark Yukawa couplings to vanish.  Thus, the up-type 
Higgsino does not give a significant contribution to any of 
$\mu \ra e \gamma$, $\mu \ra e$ conversion in nuclei 
and $\mu \ra 3e$.  

With these simplifications, the amplitudes of LFV processes are 
sensitive to just two neutralinos, mixtures of $\bino$ and $\hino_d^0$
inside the loops.  We can also neglect the contributions due 
to charginos because the only light chargino is $\hino_d^0$-like.
Hence, all types of diagrams we consider involving a chargino are 
suppressed not only by one power of muon Yukawa, but also one power of 
either the electron Yukawa, or the tiny wino content of the light chargino
at the lepton-chargino-sneutrino vertex.  This also means that sneutrino
mixing does not contribute to LFV processes, and thus
the difference in the amplitudes between left-handed and right-handed 
slepton mixing is due solely to the hypercharges and masses 
of the left-handed and right-handed charged leptons.

\subsection{$\mu \ra e \gamma$}
\label{mueg-subsec}

\begin{figure}
\subfigure[Chirality flip on the external muon line.]{
          \includegraphics[width=0.45\linewidth]{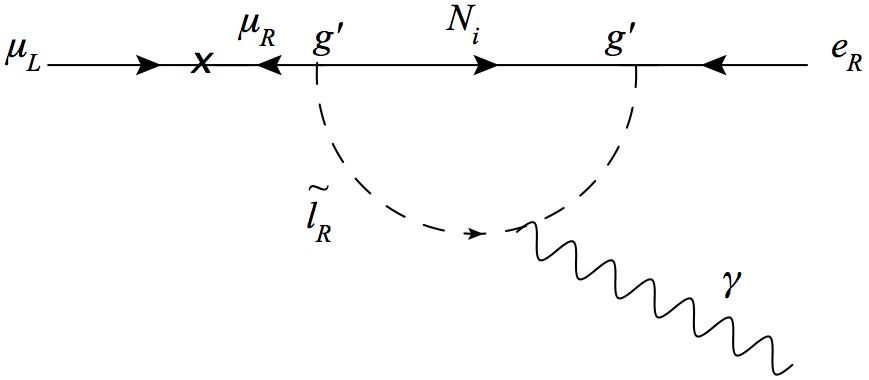}
          }
\subfigure[Chirality flip at the Yukawa vertex.]{
          \includegraphics[width=0.45\linewidth]{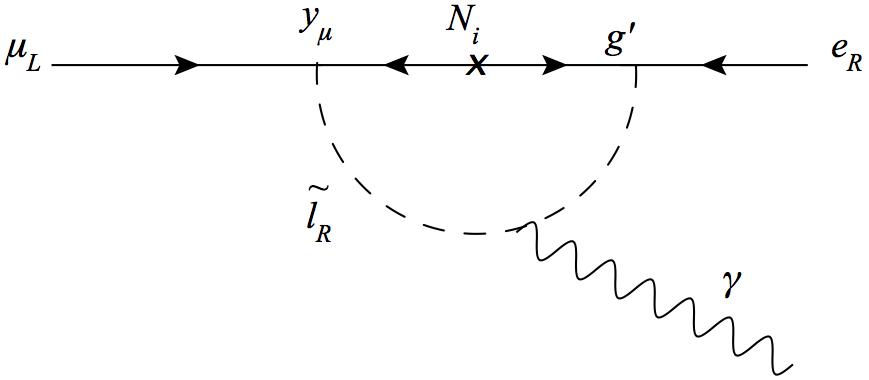}
          }	
\caption{Feynman diagrams for $\mu \ra e\gamma$ corresponding to 
the amplitudes (a) $A_{Rin1}$ and (b) $A_{Rin2}$ 
mediated by right-handed slepton flavor mixing.  
The diagrams for left-handed slepton flavor mixing are obtained 
by swapping $L \leftrightarrow R$.}
\label{mu2egammadiag}
\end{figure}

The neutrinoless muon decay $\mu \ra e\gamma$ occurs through the 
effective magnetic dipole moment operator, 
$\bar{e}\sigma_{\mu\nu}F^{\mu\nu} \mu$, 
and requires a chirality flip of fermions. 
There are no tree level operators that lead to this decay, 
and the lowest order is at one loop. In the MRSSM, there are only 
two types of contributions to the $\mu \ra e \gamma$ amplitude:
one where the chirality flip occurs on the external muon line, and 
the other where the flip occurs as a result of a muon-smuon-Higgsino
vertex proportional to the muon Yukawa coupling.  The diagrams
are shown in Fig.~\ref{mu2egammadiag}.

We calculated the amplitudes in the mass eigenstate basis of the 
sleptons and neutralinos, and as a check we derived the results 
obtained in Ref.~\cite{Hisano:1998fj} 
(replacing their $\tilde{\mu}$-$\tilde{\tau}$ mixing with 
$\tilde{e}$-$\tilde{\mu}$ mixing). 
The effective Lagrangian is
\begin{equation}
\mathscr{L}_{eff} = \frac{m_{\mu}}{2} \bar{e} \sigma_{\mu \nu} F^{\mu \nu}
(A^L_{\gamma \textrm{dip}}P_L + A^R_{\gamma \textrm{dip}}P_R)\mu.
\end{equation}
We rewrite the amplitudes 
$A^L_{\gamma \textrm{dip}}$ and $A^R_{\gamma \textrm{dip}}$, as 
\begin{eqnarray}
A^{L}_{\gamma \textrm{dip}} &=& \sum_{i=1}^2 (A_{Lin1}+A_{Lin2}) \\
A^{R}_{\gamma \textrm{dip}} &=& \sum_{i=1}^2 (A_{Rin1}+A_{Rin2}) \; ,
\end{eqnarray}
where the sum is over the $i$-th neutralinos.
The subscripts 1 and 2 denote the locations of the chirality flip
on the muon line and at the muon-slepton-gaugino vertex, respectively. 
As we shall see below, for right-handed sleptons there can be an 
accidental cancellation between amplitudes involving these diagrams.

The $\mu \ra e \gamma$ branching ratio is given by
\begin{equation}
BR(\mu \ra e \gamma) \; = \;  \frac{48\alpha\pi^3m_{\mu}^2}{G_F^2}
\bigg[|A^{L}_{\gamma \textrm{dip}}|^2+|A^{R}_{\gamma \textrm{dip}} |^2\bigg], 
\end{equation}
with the amplitudes involving a neutralino $\neut_i$ and sleptons 
$\tilde{l}_{1}$ and $\tilde{l}_{2}$ with the sleptons mass-ordered as 
$m_{\tilde{l}_{1}} < m_{\tilde{l}_{2}}$.  The amplitudes involving
right-handed sleptons are 
\begin{eqnarray}
A_{Rin1} &=& \frac{(Y^l_R)^2g'^2}{3(16\pi^2)}(O_{Li\tilde{B}})^2 \cos\theta_{\tilde{l}} \sin\theta_{\tilde{l}}\Bigg[ \frac{f_{n1}(x_{1i})}{m^2_{\tilde{l}_{R1}}} -  \frac{f_{n1}(x_{2i})}{m^2_{\tilde{l}_{R2}}} \Bigg] \label{amp1-eq},   \\ 
A_{Rin2} &=& \frac{Y^l_Rg'^2m_{\neut_i}}{2(16\pi^2)M_Z \sin\theta_w\cos\beta} O_{Ri\hino_d^0} O_{Li\bino}\cos\theta_{\tilde{l}} \sin\theta_{\tilde{l}}\Bigg[ \frac{f_{n2}(x_{1i})}{m^2_{\tilde{l}_{R1}}} -  \frac{f_{n2}(x_{2i})}{m^2_{\tilde{l}_{R2}}} \Bigg] ,
\label{amp2-eq}
\end{eqnarray}
where $A_{Rin1}$ is the amplitude that involves an external chirality flip 
of the muon and $A_{Rin2}$ involves a flip at the Higgsino vertex.  
Here $O_{Ri\hino_d^0}$ and $O_{Li\bino}$ are the 
Higgsino and bino content of $\neut_i$, respectively 
(i.e., the corresponding elements in the orthogonal matrices that 
diagonalize the gaugino mass matrix squared), and $Y^l_R = Y^{l^c} = +1$.
To lowest non-vanishing order in $M_Z$, the neutralino mixings are 
(dropping the subscripts L and R from now on):
\begin{eqnarray}
O_{1\tilde{B}}( \mu_d \ll M_1) = O_{1\hino_d^0}(\mu_d \gg M_1) &=& \frac{\cos\beta \sin\theta_w M_Z \mu_d}{M_1^2-\mu_d^2}, \label{O1-eq} \\ 
O_{2\tilde{B}}(\mu_d \gg M_1) = O_{2\hino_d^0}(\mu_d \ll M_1) &=& -\frac{\cos\beta \sin\theta_w M_Z M_1}{M_1^2-\mu_d^2}, \label{O2-eq}
\end{eqnarray}
and $O_{i(\bino,\hino_d^0)} =1$ in the appropriate limits. 
The functions $f_{nj}(x_i)$, with 
$x_{ik} = m^2_{\neut_k}/m^2_{\tilde{l}_{Ri}}$, 
with $j=1,2$, come from integrating over the loops in the diagrams:
\begin{eqnarray}
f_{n1}(x) &=& \frac{1}{2(1-x)^4}(1-6x+3x^2+2x^3-6x^2\ln{x}), \\
f_{n2}(x) &=& \frac{1}{(1-x)^3}(1-x^2+2x\ln{x}).
\end{eqnarray}
Finally, the amplitudes for the left-handed sleptons can be obtained from
the right-handed slepton results by doing the replacements
\begin{equation}
A^R_{\gamma \textrm{dip}} \ra A^L_{\gamma \textrm{dip}}
\quad \mbox{upon} \quad
(Y_R^l, m^2_{\tilde{l}_{Ri}}) \ra (Y_L^l, m^2_{\tilde{l}_{Li}}) .
\end{equation}

Inserting the results in Eqs.~(\ref{O1-eq})-(\ref{O2-eq}) 
into (\ref{amp1-eq})-(\ref{amp2-eq}), 
we see that to lowest vanishing order in $M_Z$, 
$BR(\mu \ra e \gamma)$ is independent of $\tan\beta$. 
We can also see explicitly that when the two slepton masses 
are degenerate, the branching ratio vanishes, as expected from 
the super GIM mechanism.

As an aside, it is also straightforward to see what happens
to the results when the mass hierarchy between the slepton and
the neutralino are inverted.  The loop functions satisfy the
identities,
\begin{eqnarray}
f_{n1}(x) + f_{n1}\bigg(\frac{1}{x}\bigg) &=& \frac{1}{2}, \\
xf_{n2}(x) - f_{n2}\bigg(\frac{1}{x}\bigg) &=& 0.
\end{eqnarray}

We are now in a position to discuss the amplitudes in various limits. 
In the bino-like limit $M_1 \ll \mu_d$, one sees that 
$A_{R1n1}$ dominates, as $A_{Rin2}$ is of order $M_1/\mu_d$. 

When $\neut_1$ becomes $\hino^0_d$-like, there is a cancellation 
between the amplitudes involving a chirality flip on the 
external muon line, and the one with the flip occurring at the 
muon Yukawa vertex. The dominant diagram in the $\tilde{B}$-like case, 
$A_{R1n1}$, is now suppressed by $\mu_d^2/M_1^2$, the same 
suppression factor appears $A_{R1n2}$. So the dominant amplitudes 
come from the diagrams involving a $\tilde{B}$-like neutralino exchange. 
Note that $A_{R2n2}$ has an opposite sign compared to $A_{R2n1}$ 
and the total amplitude can vanish for some choice of parameters. 

\begin{figure}
\shortstack[c]{$M_1$ \\ $\mbox{[GeV]}$} 
\subfigure[$\mu_d = 100$ GeV]{
          \includegraphics[width=0.45\linewidth]{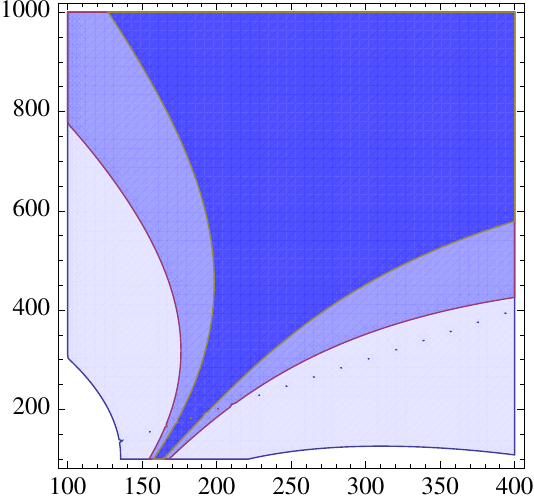}
          }
\subfigure[$\mu_d = 200$ GeV]{
          \includegraphics[width=0.45\linewidth]{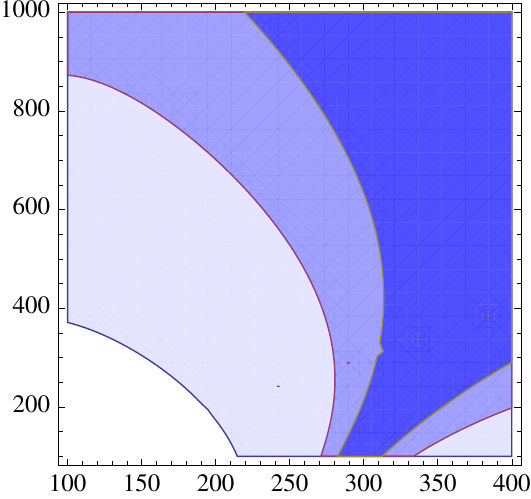}
          } \\
\shortstack[c]{$M_1$ \\ $\mbox{[GeV]}$} 
\subfigure[$\mu_d = 300$ GeV]{
          \includegraphics[width=0.45\linewidth]{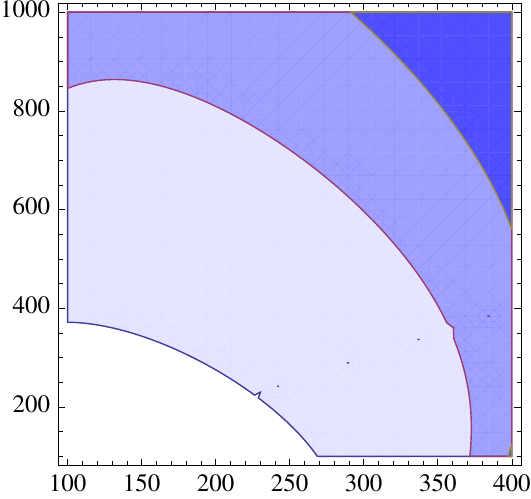}
          }
\subfigure[$\mu_d = 400$ GeV]{
          \includegraphics[width=0.45\linewidth]{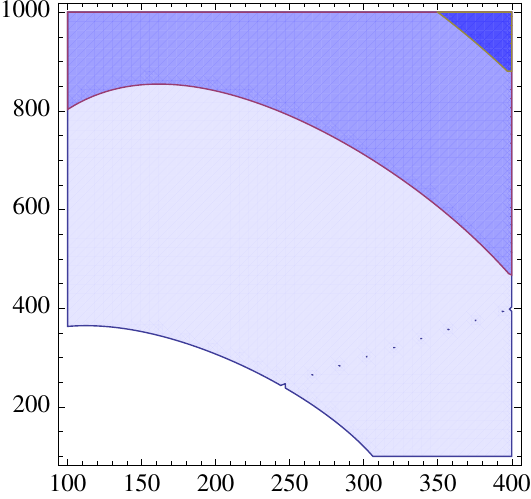}
          } \\ 
\hspace*{0.24\textwidth} $m_{\tilde{l}_1}$~~[GeV]
\hspace*{0.35\textwidth} $m_{\tilde{l}_1}$~~[GeV]
\caption{Regions in parameter space (shaded) that satisfy the 
$\mu \ra e \gamma$ bound for right-handed slepton mixing. 
The mass of the heavier slepton is 
set to $1.5 m_{\tilde{l}_1}$.  From light to dark, the shaded areas 
denote mixing with $\sin2\theta_{\tilde{l}}=0.1, 0.5$ and $1$, 
respectively. The funnel regions in the plots with $\mu_d = 100, 200$~GeV 
is caused by the cancellation between the amplitudes involving the 
bino-like and the $\hino_d^0$-like neutralinos.}
\label{mu2eright}
\end{figure}

In Figs.~\ref{mu2eright}(a)-\ref{mu2eright}(d), we show the allowed
regions in MRSSM parameter space with right-handed slepton mixing
that satisfy the bound $BR(\mu \ra e\gamma) < 1.2 \times 10^{-11}$
\cite{Ahmed:2001eh,Adam:2009ci}.

\begin{figure}
\shortstack[c]{$M_1$ \\ $\mbox{[GeV]}$}
\subfigure[$\mu_d = 100$ GeV]{
          \includegraphics[width=0.45\linewidth]{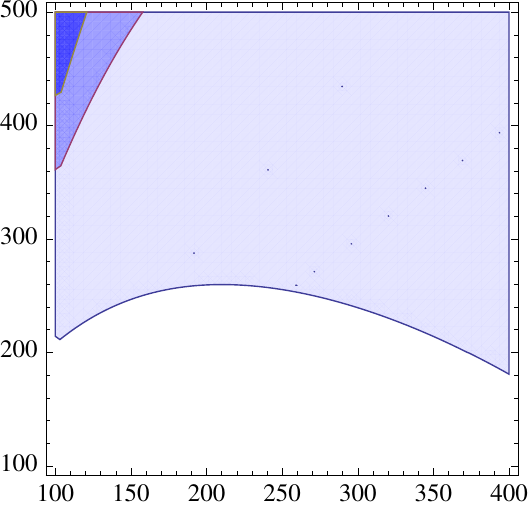}
          }
\subfigure[$\mu_d = 200$ GeV]{
          \includegraphics[width=0.45\linewidth]{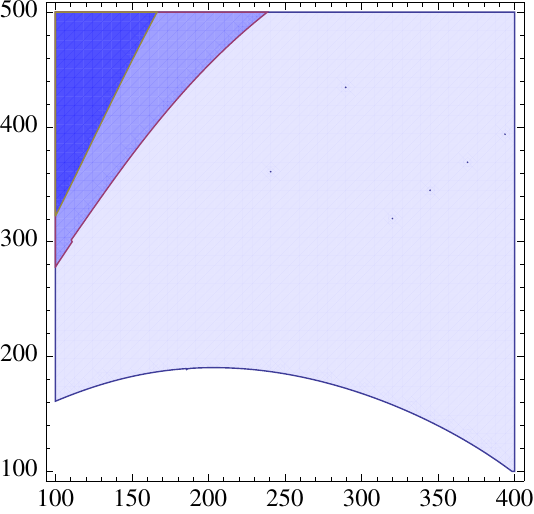}
          } \\	
\shortstack[c]{$M_1$ \\ $\mbox{[GeV]}$}
\subfigure[$\mu_d = 300$ GeV]{
          \includegraphics[width=0.45\linewidth]{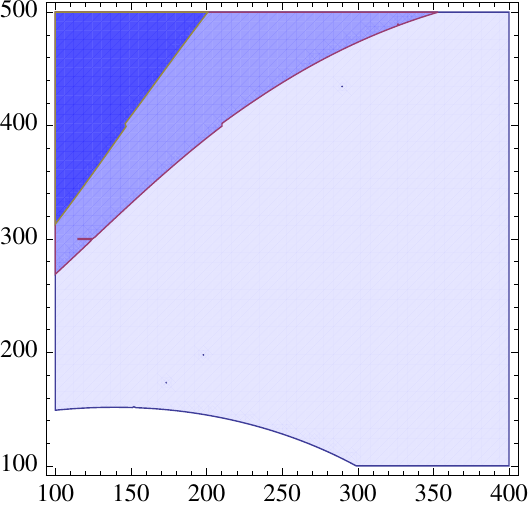}
          }
\subfigure[$\mu_d = 400$ GeV]{
          \includegraphics[width=0.45\linewidth]{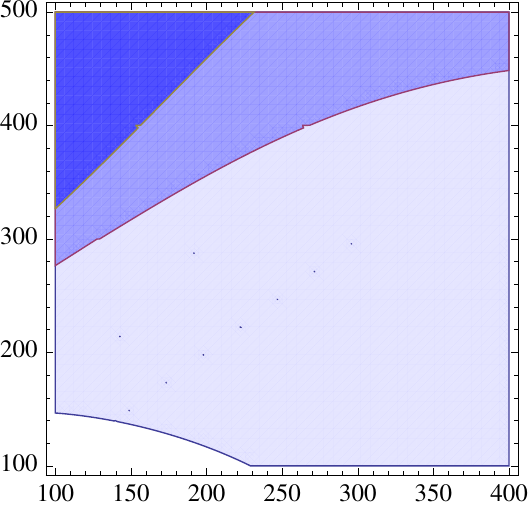}
          } \\ 
\hspace*{0.24\textwidth} $m_{\tilde{l}_1}$~~[GeV]
\hspace*{0.35\textwidth} $m_{\tilde{l}_1}$~~[GeV]
\caption{Same as Fig.~\ref{mu2eright} but for left-handed slepton mixing. 
We have restricted $M_1 < 500$~GeV since contributions from wino-like 
charginos not been included (see Sec.~\ref{simp-sec} for a discussion).}
\label{mu2eleft}
\end{figure}

The situation is drastically different in the case of left-handed 
slepton mixing. The hypercharge of the left-handed leptons 
($Y_L^l = -1/2$), has an opposite sign to the right-handed 
lepton hypercharge, and so the amplitudes interfere constructively, 
instead of destructively as in the case of right-handed slepton mixing. 
This leads to a more severe bound on the allowable regions in 
parameter space for left-handed slepton mixing. This is shown in 
Figs.~\ref{mu2eleft}(a)-\ref{mu2eleft}(d).

\subsection{$\mu \ra e$ conversion in a nucleus}
\label{muec-subsec}

The conversion of a muon into an electron can give a qualitatively distinct
bound on $\mu \leftrightarrow e$ slepton mixing because there are 
several types of operators beyond those that contribute to 
$\mu \ra e \gamma$.  We discuss the operators for $\mu \ra e$ 
conversion, one-by-one, in this section.

The $\mu \ra e$ conversion amplitude is dominated by coherent processes,
and so we only took the quark vector currents into account. 
The operators that contribute to the incoherent terms, 
$\bar{q}\gamma^5q$, $\bar{q} \gamma^{\mu} \gamma^5q$, 
and $\bar{q} \sigma^{\mu \nu} q$ have been neglected. 
This leaves us with the scalar and vector current, 
$\bar{q}q$ and $\bar{q}\gamma^{\mu}q$, respectively
\cite{Hisano:1995cp}.

The only diagram that can contribute to a scalar quark current is 
the box diagram. Without left-right mixing of sleptons in the MRSSM, 
the dominant term, with bino couplings 
at each vertex, contains no chirality flip of the quarks, 
and is therefore a vector current. We also take the quark current to be 
non-relativistic to simplify the calculation involving the 
magnetic dipole term. Thus, the amplitude for $\mu \ra e$ conversion 
is well approximated, for our purposes, by only taking quark vector 
currents into account.

\begin{figure}
\centering
\includegraphics[width=0.45\linewidth]{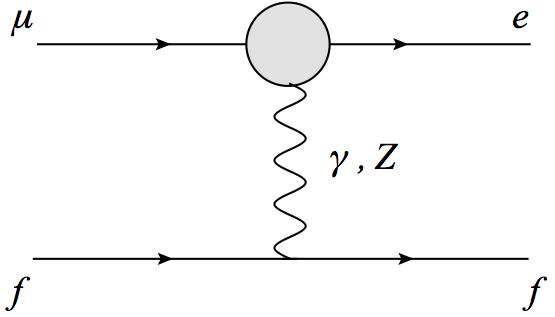}
\caption{Schematic diagram illustrating the set of penguin 
contributions to $\mu \ra e$ conversion (for $f=q$) as well as 
$\mu \ra 3e$ (for $f=e$).  The blob in the figure arises from
both charge radius subdiagrams shown in Fig.~\ref{effvertex}, 
as well as Z penguin subdiagrams, the dominant ones shown in
Fig.~\ref{Zpenguin}.}
\label{generalpenguin}
\end{figure}

\begin{figure}
\centering
\includegraphics[width=0.45\linewidth]{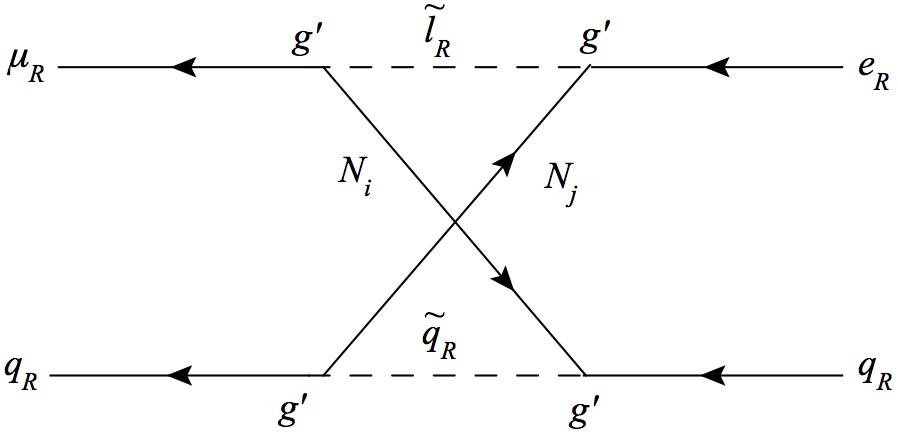}
\caption{The box Feynman diagram for $\mu \ra e$ conversion. 
Due to the conservation of $R$-charges, the chirality of 
the squarks must be the ones shown in the diagrams.
Only right-handed slepton flavor mixing diagrams are shown, while
left-handed slepton flavor mixing diagrams are obtained 
by swapping $L \leftrightarrow R$ everywhere.}
\label{BoxI}
\end{figure}

The diagrams we consider are the photon penguin, the $Z$ penguin, 
and the box diagram shown in 
Figs.~\ref{generalpenguin},\ref{effvertex},\ref{Zpenguin},\ref{BoxI}.
We only take the dominant terms of the box and 
the $Z$ penguin amplitude into account:  that is, the terms involving 
the bino coupling at each vertex which does not contain any chirality flips 
of the external fermions.  The effective Lagrangian at the parton level 
can be written as \cite{Hisano:1995cp}
\begin{eqnarray}
\mathscr{L}_{eff} &=&  \sum_{q=u,d}-Q_qe^2 \bar{e}\bigg[\gamma^{\mu}(A_{\gamma}^L P_L + A_{\gamma}^R P_R)+\frac{m_{\mu}}{k^2}i\sigma^{\mu \nu}k_{\nu}(A^L_{\gamma \textrm{dip}}P_L+A^R_{\gamma \textrm{dip}}P_R)\bigg]\mu  \bar{q} \gamma_{\mu} q \nonumber\\
 & &{} + e^2\sum_{q=u,d} \bar{e} \gamma^{\mu}[(A_Z^L+A_{ \textrm{box}}^{qL}) P_L + (A_Z^R+A_{\textrm{box}}^{qR}) P_R]\mu \bar{q}\gamma_{\mu} q,
\end{eqnarray}
where $Q_q$ is the quark electric charge, 
$k^2 \sim -m_{\mu}^2$ is the momentum transfer, 
$A_{\gamma ,Z}^{L,R}$ and $A^{L,R}_{\gamma \textrm{dip}, Z}$ correspond 
to the $\gamma$-penguin and $Z$-penguin, respectively, 
and $A_{ \textrm{box}}^{q(L,R)}$ corresponds to the box diagram.

The most severe upper bound to date is on the conversion rate ratio 
with a gold nucleus 
$BR(\mu \ra e)_{Au} \equiv 
\Gamma(\mu^- \, \mathrm{Au} \ra e^- \, \mathrm{Au})/
\Gamma(\mu^- \, \mathrm{Au})_{\mathrm{capture}} < 7 \times 10^{-13}$ 
from SINDRUM II \cite{Bertl:2006up}.
Because of the large number 
of protons in the gold nucleus, the distortion to the muon wave function 
from a plane wave must be taken into account when evaluating 
the overlap between the muon and nucleus wavefunctions. 
This has been done in Ref.~\cite{Kitano:2002mt}, and we will use their 
overlap integrals, with the neutron density determined from 
pionic atom experiments (method 2 in \cite{Kitano:2002mt}). 
Other nuclei could also be of interest, particularly as a way
to distinguish different models \cite{Cirigliano:2009bz}.
The conversion rate is
\begin{equation}
\Gamma_{\mu \ra e} \; = \; 4m_{\mu}^5e^4|\mathscr{A}_{\gamma \textrm{dip}}^L + \mathscr{A}_{\gamma}^R + \mathscr{A}_{box}^R + \mathscr{A}_{Z}^R|^2  + (L \leftrightarrow R),
\label{mu2econvrate}
\end{equation}
where,
\begin{eqnarray}
\mathscr{A}_{\gamma \textrm{dip}}^L &=& -\frac{1}{8e}A^L_{\gamma \textrm{dip}} D, \\
\mathscr{A}_{\gamma}^R &=& A_{\gamma}^R V^{(p)}, \\
\mathscr{A}_{box}^R &=&  - (2A_{\textrm{box}}^{uR}+A_{\textrm{box}}^{dR})V^{(p)} - (A_{\textrm{box}}^{uR}+2A_{\textrm{box}}^{dR})V^{(n)}, \\
\mathscr{A}_{Z}^R &=&  [(2Z_u+Z_d)V^{(p)} + (2Z_d+Z_u)V^{(n)}]A_Z^R,
\end{eqnarray}
where 
$Z_q = (Z_{q_R} + Z_{q_L})/2$, with 
$Z_{q_{(L,R)}} = I^{q}_{L,R} - Q \sin^2 \theta_w$, 
$I^{u}_L = 1/2$, $I^{d}_L = -1/2$ for up and down type quarks, 
and $I^q_R = 0$.
The first term in Eq.~(\ref{mu2econvrate}), proportional to 
$|\mathscr{A}_{\gamma \textrm{dip}}^L + \mathscr{A}_{\gamma}^R + \mathscr{A}_{box}^R + \mathscr{A}_{Z}^R|^2$, corresponds to 
slepton mixing in the right-handed sector, while the second term
proportional to 
$|\mathscr{A}_{\gamma \textrm{dip}}^R + \mathscr{A}_{\gamma}^L + \mathscr{A}_{box}^L + \mathscr{A}_{Z}^L|^2$, corresponds to 
to slepton mixing in the left-handed sector. 
The coefficients $D$ and $V^{(p,n)}$ are to the overlap integrals 
of the muon and the nucleus for the leptonic dipole and vector 
(proton, neutron) operators.  We used, for a gold nucleus, $D = 0.167$, 
$V^{(p)} = 0.0859$, $V^{(n)} = 0.108$ from Ref.~\cite{Kitano:2002mt}.

Now we will discuss each 
diagram below. We will present the results for both left- and 
right-handed slepton mixing. But, for simplicity, we will only discuss 
the case of right-handed slepton mixing explicitly. 
The amplitudes corresponding to left-handed slepton mixing can be 
obtained from the right-handed ones by replacing the appropriate 
hypercharges and slepton masses. Note that for the $Z$-penguin, 
there is also an additional minus sign after the replacement 
of hypercharges and slepton masses.

\subsubsection{Charge radius}

\begin{figure}
\subfigure[]{
            \includegraphics[width=0.3\linewidth]{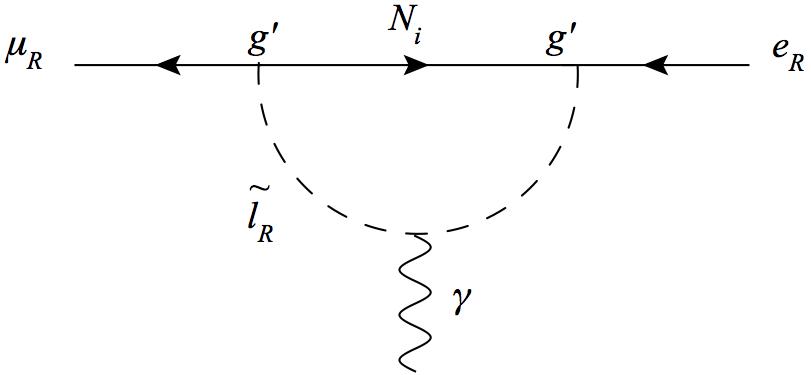}
            }
\subfigure[]{
            \includegraphics[width=0.3\linewidth]{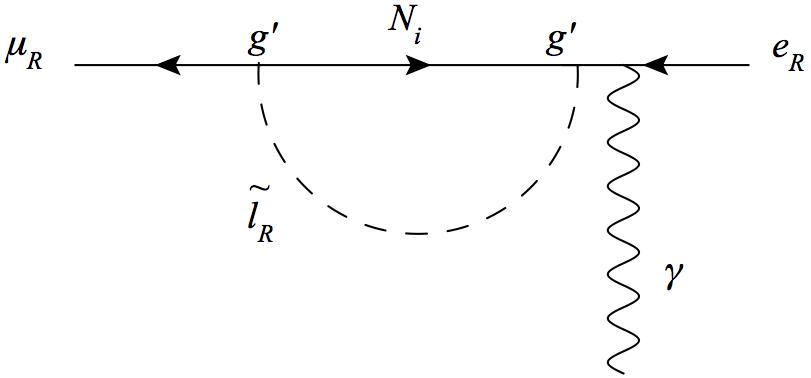}
            }	
\subfigure[]{
            \includegraphics[width=0.3\linewidth]{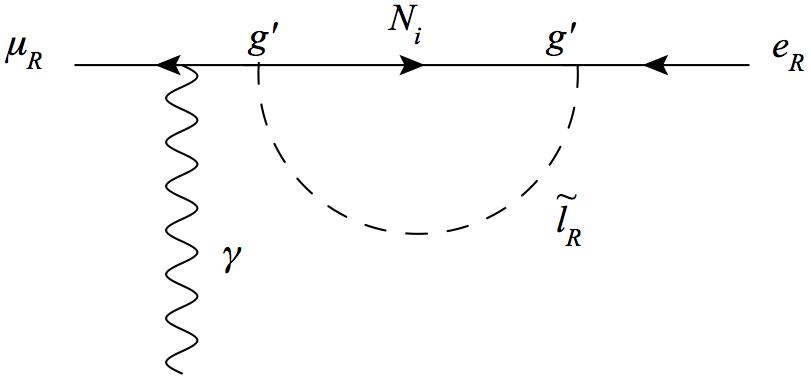}
            }	
\caption{Contributions to the effective vertex from the charge radius
operator. 
Graph (c) is suppressed by a factor of $m_e^2/m_{\mu}^2$ compared to (b), 
and can be ignored in the limit of vanishing electron mass. 
Also in this limit, graph (b) exactly cancels graph (a) for 
vanishing photon momentum, satisfying the Ward identity.
Only right-handed slepton flavor mixing diagrams are shown, while
left-handed slepton flavor mixing diagrams are obtained 
by swapping $L \leftrightarrow R$.}
\label{effvertex}
\end{figure}

The charge radius amplitude $A_{\gamma}^{L,R}$ comes from 
the $\gamma$-penguin, without a chirality flip of the leptons. 
The dominant term is the one involving the $\tilde{B}$-like 
neutralino in the loop, with $\tilde{B}$ coupling at each vertex 
connecting a lepton. The other terms are suppressed either by the 
muon Yukawa or by two powers of the small bino content in the 
$\hino_d^0$-like neutralino. The contributions to the 
effective vertex of the charge radius is shown in Fig.~\ref{effvertex}. 
Summing over these contributions give\footnote{We have checked that, 
even when $\mu_d = M_1$, the value given by this expression differs 
to the exact one by $\lesssim$ 1\%. So this expression is valid 
over all ranges of $M_1$ and $\mu_d$. The discrepancy comes from 
the small mass splitting of the neutralinos when the gaugino 
and Higgsino masses are degenerate. We have used the exact expression 
in our numerical analysis.},
\begin{equation}
A_{\gamma}^{R}=\frac{g'^2 (Y^l_{R})^2}{576\pi^2 } \frac{\sin 2\theta_{\tilde{l}}}{m_{\tilde{l}_1}^2} f_{\gamma}\bigg(\frac{M_1^2}{m_{\tilde{l}_1}^2}\bigg)- (m_{\tilde{l}_1} \ra m_{\tilde{l}_2}),
\end{equation}
with
\begin{equation}
f_{\gamma}(x) = \frac{1}{1-x^4} (2-9x+18x^2-11x^3+6x^3\ln x).
\end{equation}
%

\subsubsection{Magnetic dipole}

The magnetic dipole amplitude $A^{L,R}_{\gamma \textrm{dip}}$ is 
the one that appears in $\mu \ra e \gamma$, which was discussed 
in detail in the last section. For right-handed slepton mixing, 
the amplitude of the dipole term is smaller than the charge radius term, 
$A^{L,R}_{\gamma}$, due to the destructive interference between 
amplitudes involving chirality flips at different locations 
in the diagram. The situation reverses in the case of left-handed 
slepton mixing, where both terms contributes and the magnitude 
becomes larger than the charge radius term.

\subsubsection{Z-penguin}

\begin{figure}
\subfigure[]{
            \includegraphics[width=0.45\linewidth]{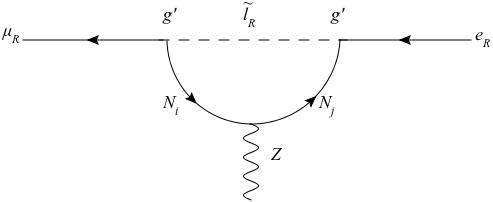}
            }
\subfigure[]{
            \includegraphics[width=0.45\linewidth]{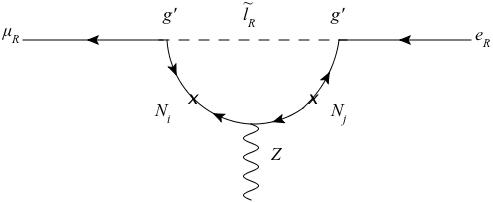}
            }	
\caption{Contributions to the effective vertex from the $Z$ penguin. 
Diagram (a) gives the term 
proportional to $f_Z$ in which the $Z$ boson couples to the $R$-partner 
of the down type Higgsino, $\psi_{\hino_d^0}$, and (b) gives the term 
proportional to $g_Z$, with $Z$ coupling to $\hino_d^0$.
Only right-handed slepton flavor mixing diagrams are shown, while
left-handed slepton flavor mixing diagrams are obtained 
by swapping $L \leftrightarrow R$.}
\label{Zpenguin}
\end{figure}

The $Z$-penguin contribution contains diagrams in Fig.~\ref{effvertex}, 
with the photon replaced by the $Z$ boson. The contribution coming 
from this set of diagrams is suppressed by $O(m_{\mu}^2/M_Z^2)$ 
compared to the charge radius so is negligible. 
Then, the dominant term is the one involving a Higgsino-Higgsino-$Z$ vertex, 
shown in Fig.~\ref{Zpenguin}.

We find that the $Z$-penguin is sub-dominant in a large region 
of the parameter space. The $Z$-penguin is the only 
amplitude that is sensitive to $\tan \beta$, and in the limit 
$M_Z \ll M_{\neut}$, it scales as $\cos^2 \beta$. 
The $Z$-penguin amplitude is
\begin{equation}
A_Z^R = \frac{(Y^l_R)^2g'^2}{64 \pi^2}\frac{\sin2\theta_{\tilde{l}}}{M_Z^2 \sin^2\theta_w \cos^2\theta_w} \sum_{i,j=1}^2 \omega_{ij},
\end{equation}
where
\begin{equation}
\omega_{ij} = O_{Li1}O_{Lj1}\bigg[O_{Li2}O_{Lj2} f_Z\bigg(\frac{M_{\neut_i}^2}{m_{\tilde{l}_1}^2},\frac{M_{\neut_j}^2}{m_{\tilde{l}_1}^2}\bigg) - 2O_{Ri2}O_{Rj2}g_Z\bigg(\frac{M_{\neut_i}^2}{m_{\tilde{l}_1}^2},\frac{M_{\neut_j}^2}{m_{\tilde{l}_1}^2}\bigg)\bigg] -(m_{\tilde{l}_1} \ra m_{\tilde{l}_2}).
\end{equation}
The functions $f_Z(x_i,x_j)$ and $g_Z(x_i,x_j)$ are\footnote{Note that 
the function $f_Z$ appears to contain a log term that is asymmetric 
in the two neutralino lines in the loop, not as one would expect. 
But remember that this log term is subtracted by one containing 
the heavier slepton mass, and the final result is symmetric in the 
neutralinos and anti-symmetric in the sleptons, as expected.}
\begin{eqnarray}
f_Z(x_i,x_j) &=& \ln x_i + \frac{1}{x_i-x_j}\bigg[ \frac{x_i^2 \ln x_i}{1-x_i}- \frac{x_j^2 \ln x_j}{1-x_j} \bigg], \\
g_Z(x_i, x_j) & = & \frac{\sqrt{x_i x_j}}{x_i - x_j}\bigg[ \frac{x_i \ln x_i}{1-x_i}- \frac{x_j \ln x_j}{1-x_j} \bigg].
\end{eqnarray}

Note that the $Z$-penguin effective vertex does not explicitly 
depend on $1/M^2_{\textrm{SUSY}}$ as in the case of all other amplitudes. 
This corresponds to an operator of dimension-4.  
This is perfectly fine, because the weak symmetry is broken, 
so the weak current is not conserved. However, it is required that 
in the limit of unbroken electroweak symmetry, this effective 
vertex vanishes. This is easy to check in the limit $M_Z \ra 0$. 
In this limit, the neutralinos we consider do not mix 
[c.f., Eq.~(\ref{mixingmatrix})].  But the amplitude for the 
$Z$-penguin contain at least two powers of the neutralino mixing 
matrix elements, regardless of whether it is bino-like or Higgsino-like. 
Therefore this operator vanishes in the limit $M_Z \ra 0$, 
when the electroweak symmetry is unbroken.

For left-handed sleptons, the $Z$ amplitude can be obtained 
by replacing the appropriate hypercharges and slepton masses, 
as well as an additional factor of $(-1)$. This sign change arises 
from the $\neut\neut Z$ coupling, in contrast to $\neut^c\neut^cZ$ 
in the case of right-handed sleptons. 

\subsubsection{Box diagram}

\begin{figure}
\centering
\includegraphics[width=0.5\linewidth]{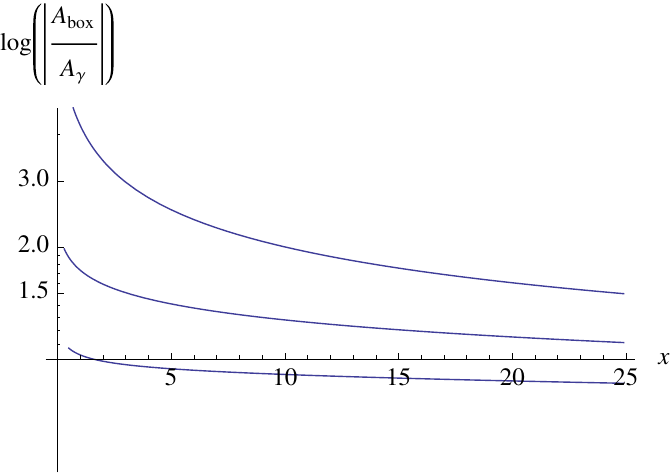}
\caption{A plot of the right hand side of Eq.~(\ref{AboxonAgammaR}), 
$19j_4(x,x,y)/f_{\gamma}(x)$, where $x = M_1^2/m_{\tilde{l}_1}^2$ and 
$y = m_{\tilde{q}}^2/m_{\tilde{l}_1}^2$. 
The contours are $y=1,10,25$ from top to bottom.  
The box amplitude is larger than the electromagnetic term when 
the contour is above the $x$-axis.}
\label{AboxonAgammaRplot}
\end{figure}

For the box diagram, the dominant term is the one containing 
bino couplings at all four vertices,
\begin{equation}
A_{box}^{qR} = \frac{(Y^l_R)^2g'^4 \sin2\theta_{\tilde{l}}}{64 \pi^2 e^2 m_{\tilde{l}_1}^2}\bigg[ (Y^q_R)^2j_4\bigg(\frac{M_1^2}{m_{\tilde{l}_1}^2},\frac{M_1^2}{m_{\tilde{l}_1}^2}, \frac{m_{{\tilde{q}_R}}^2}{m_{\tilde{l}_1}^2}\bigg)\bigg] - (m_{\tilde{l}_1} \ra m_{\tilde{l}_2}),
\end{equation}
where
\begin{equation}
j_4(x_i,x_j,y) = \frac{x_i^2 \ln x_i}{(1-x_i)(x_i-x_j)(x_i-y)} - \frac{x_j^2 \ln x_j}{(1-x_j)(x_i-x_j)(x_j-y)} + \frac{y^2 \ln  y}{(1-y)(x_i-y)(x_j-y)}.
\end{equation}
We can compare the box amplitude with $A_{\gamma}^{L,R}$ 
by approximating $V^{(p)} \simeq V^{(n)}$, giving
\begin{equation}
\bigg| \frac{A_{box}^R}{A_{\gamma}^{R}} \bigg| \; = \; \frac{9 (g')^2}{e^2} \frac{j_4(x,x,y)}{f_{\gamma}(x)}[3(Y_R^d)^2+3(Y_R^u)^2] \; \simeq \; 19 \frac{j_4(x,x,y)}{f_{\gamma}(x)},
\label{AboxonAgammaR}
\end{equation}
where $x = M_1^2/m_{\tilde{l}_1}^2$ and 
$y = m_{\tilde{q}}^2/m_{\tilde{l}_1}^2$. 
The right hand side is plotted in Fig.~\ref{AboxonAgammaRplot}. 
We can see that the box can give a large contribution the 
total amplitude when the squarks are not far heavier than the sleptons. 

\subsubsection{Numerical Results}

\begin{figure}
\subfigure[$\mu_d = 100$ GeV, $m_{\tilde{l}}=200$ GeV]{
          \includegraphics[width=0.47\linewidth]{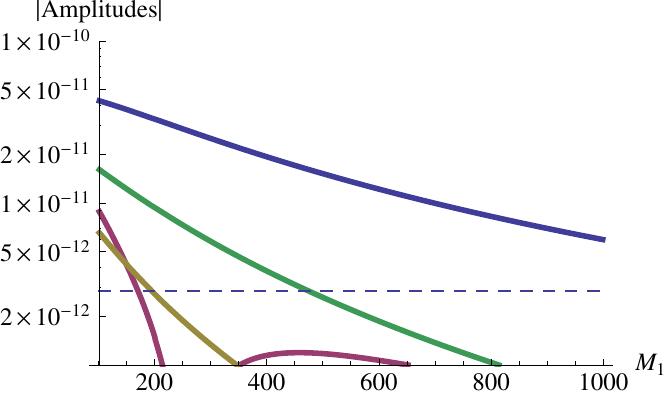}
          }
\subfigure[$\mu_d = 200$ GeV, $m_{\tilde{l}}=200$ GeV]{
          \includegraphics[width=0.47\linewidth]{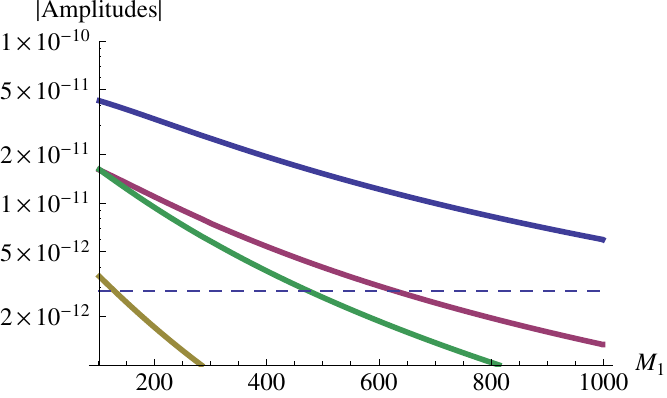}
          } \\
\subfigure[$M_1=\mu_d = 100$ GeV]{
          \includegraphics[width=0.47\linewidth]{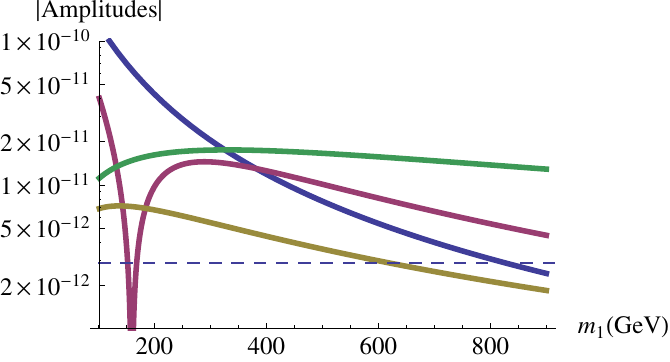}
          }
\subfigure[$M_1=\mu_d = 200$ GeV]{
          \includegraphics[width=0.47\linewidth]{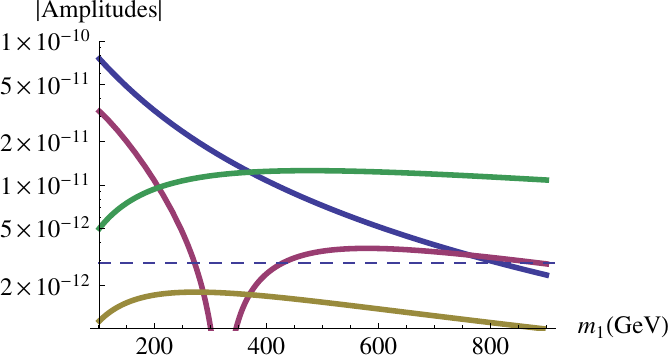}
          }		
\caption{The magnitudes of various amplitudes at maximal mixing of 
right-handed sleptons with degenerate squark masses of $1$~TeV 
(i.e., the terms in Eq.~(\ref{mu2econvrate}) before taking the square). 
The contours are, 
$\mathscr{A}_{\gamma}^R$ (blue), 
$\mathscr{A}_{box}^R$ (green), 
$|\mathscr{A}_{\gamma \textrm{dip}}^R|$ (red), and 
$-\mathscr{A}_Z^R$ (brown). 
The dashed line corresponds to the bound on $\mu \ra e$ conversion 
as if only one amplitude were contributing. One can see that there 
are regimes where only the box and the charge radius amplitudes 
contribute significantly [subfigures (a) and (b), especially in the 
high $M_1$ regions in these figures], and where all four amplitudes 
contribute significantly [subfigure (c)]. 
In subfigure (d), the magnetic dipole amplitude 
reaches zero near $m_{\tilde{l}_1}\sim 330$~GeV\@. 
This coincides with the ``funnel'' region in the parameter space plot 
for $\mu \ra e \gamma$, Fig.~\ref{mu2eright}(b).}
\label{ampplot}
\end{figure}

\begin{figure}
\subfigure[$\mu_d = 100$ GeV, $m_{\tilde{l}}=200$ GeV]{
          \includegraphics[width=0.47\linewidth]{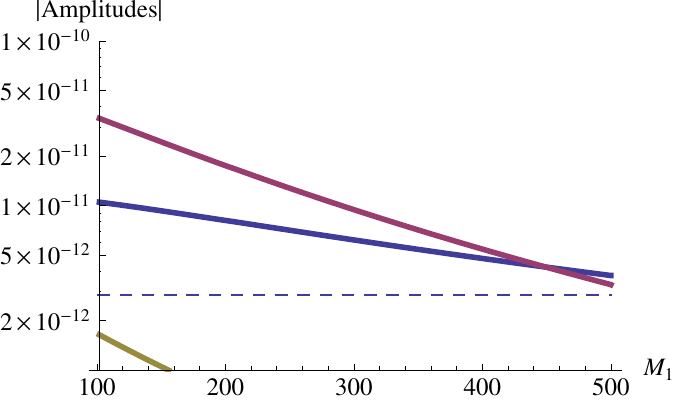}
          }
\subfigure[$\mu_d = 200$ GeV, $m_{\tilde{l}}=200$ GeV]{
          \includegraphics[width=0.47\linewidth]{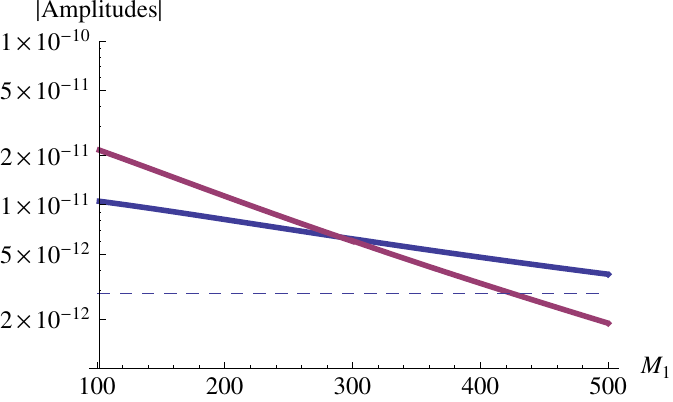}
          } \\	
\subfigure[$M_1=\mu_d = 100$ GeV]{
          \includegraphics[width=0.47\linewidth]{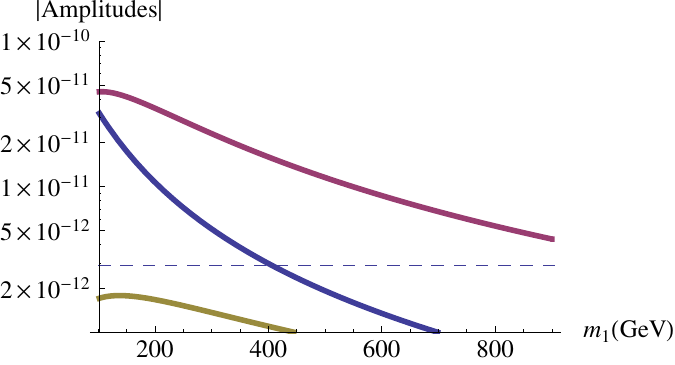}
          }
\subfigure[$M_1=\mu_d = 200$ GeV]{
          \includegraphics[width=0.47\linewidth]{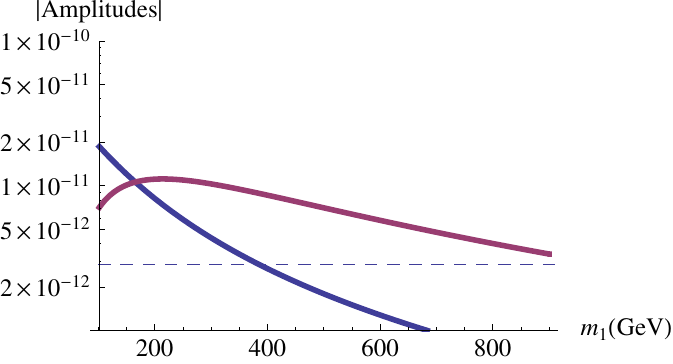}
          }		
\caption{Same as Fig.~\ref{ampplot} but with left-handed 
slepton mixing instead. The contours are, 
$\mathscr{A}_{\gamma}^L$ (blue), 
$-\mathscr{A}_{box}^L$ (green), 
$-\mathscr{A}_{\gamma \textrm{dip}}^L$ (red), and 
$\mathscr{A}_Z^L$ (brown). 
The magnetic dipole and the charge radius amplitudes interfere 
destructively with each other, opening up a large region in the 
parameter space that satisfies $\mu \ra e$ conversion. 
This forms the funnel regions in Fig.~\ref{mu2eL}.}
\label{ampplotL}
\end{figure}

We took $\tan \beta = 3$ for our analysis.  The amplitudes 
contributing to $\mu \to e$ conversion in gold are shown 
in Fig.~\ref{ampplot} for right-handed slepton mixing, 
and in Fig.~\ref{ampplotL} for left-handed slepton mixing.  
The slepton mixing angles are taken to be maximal. 
For comparison, we also drew the line where the experimental bound 
on the amplitude would be, as if only one amplitude were contributing 
to the conversion rate.

For right-handed sleptons, either the charge radius or the box diagram
dominate over other contributions.  Each of these amplitudes exceeds 
the bound alone and they interfere constructively with each other. 
Therefore, maximal right-handed slepton mixing is excluded throughout 
the parameter space we explore.  The magnetic dipole destructively 
interferes with the box and the charge radius diagrams, 
at small slepton masses before the magnetic dipole vanishes. 
However, this cancellation is insufficient to bring the amplitudes 
below the bound.

In the left-handed slepton mixing case, the box diagram is suppressed 
by the left-handed quark hypercharge, and is much smaller.  
Also, the two largest amplitudes, the charge radius and the magnetic dipole, 
destructively interfere with each other, resulting in the funnel region 
shown in Fig.~\ref{mu2eL}.

\begin{figure}
\shortstack[c]{$M_1$ \\ $\mbox{[GeV]}$}
\subfigure[$\mu_d = 100$ GeV]{
          \includegraphics[width=0.45\linewidth]{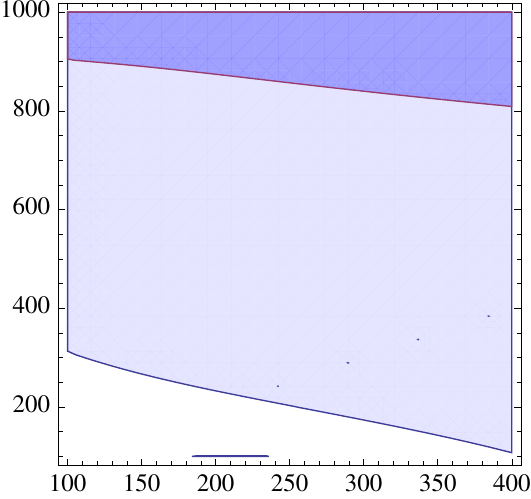}
          }
\subfigure[$\mu_d = 200$ GeV]{
          \includegraphics[width=0.45\linewidth]{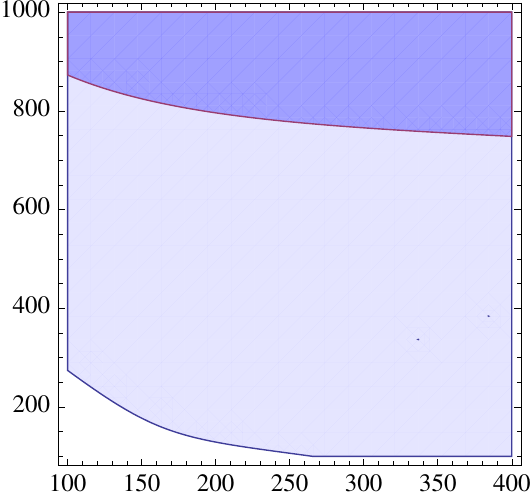}
          } \\
\shortstack[c]{$M_1$ \\ $\mbox{[GeV]}$}
\subfigure[$\mu_d = 300$ GeV]{
          \includegraphics[width=0.45\linewidth]{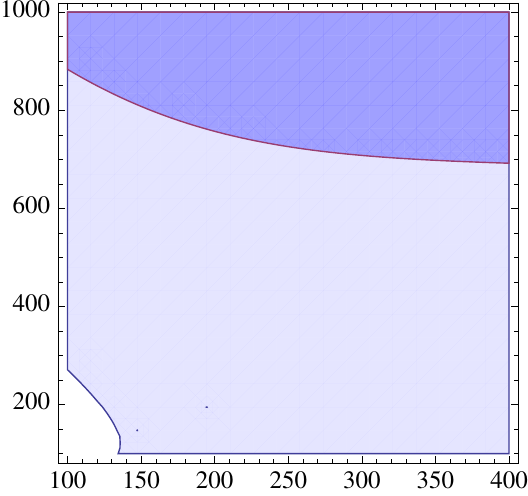}
          }	
\subfigure[$\mu_d = 400$ GeV]{
          \includegraphics[width=0.45\linewidth]{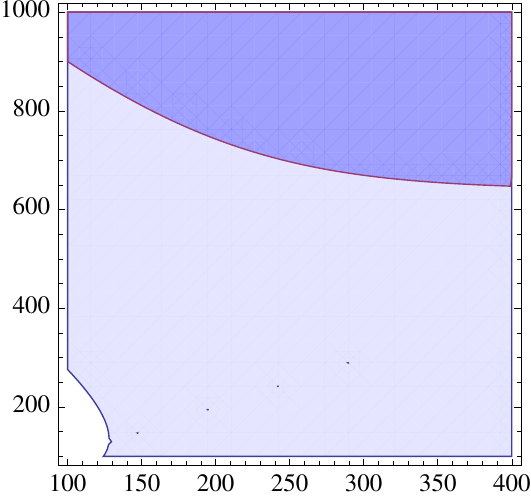}
          } \\ 
\hspace*{0.24\textwidth} $m_{\tilde{l}_1}$~~[GeV]
\hspace*{0.35\textwidth} $m_{\tilde{l}_1}$~~[GeV]
\caption{Allowable regions for $\mu \ra e$ conversion in a 
gold nucleus with right-handed slepton mixing. 
From light to dark, the shaded areas denote mixing 
with $\sin2\theta_{\tilde{l}}=0.1, 0.5$ respectively. 
The squark masses are set to be degenerate at $1$~TeV\@. 
Note that this completely rules out maximal mixing for 
right-handed sleptons in the sub-TeV range.}
\label{mu2eR}
\end{figure}

\begin{figure}
\shortstack[c]{$M_1$ \\ $\mbox{[GeV]}$}
\subfigure[$\mu_d = 100$ GeV]{
          \includegraphics[width=0.45\linewidth]{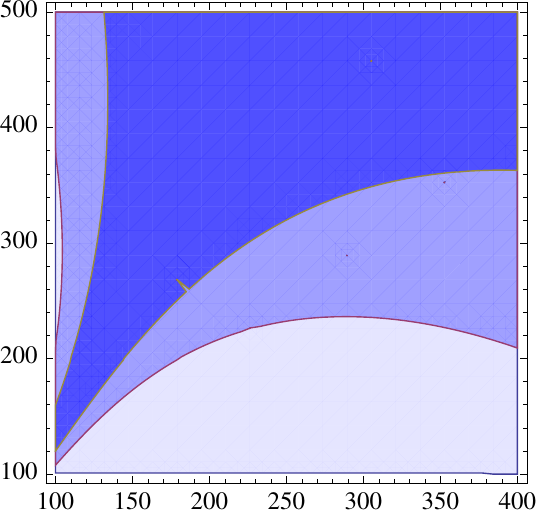}
          }
\subfigure[$\mu_d = 200$ GeV]{
          \includegraphics[width=0.45\linewidth]{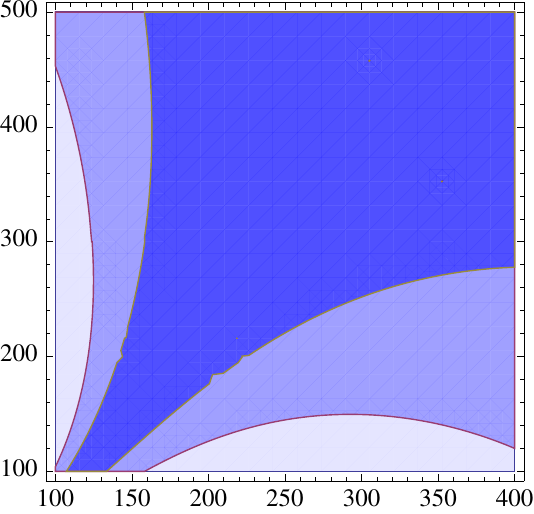}
          } \\
\shortstack[c]{$M_1$ \\ $\mbox{[GeV]}$}
\subfigure[$\mu_d = 300$ GeV]{
          \includegraphics[width=0.45\linewidth]{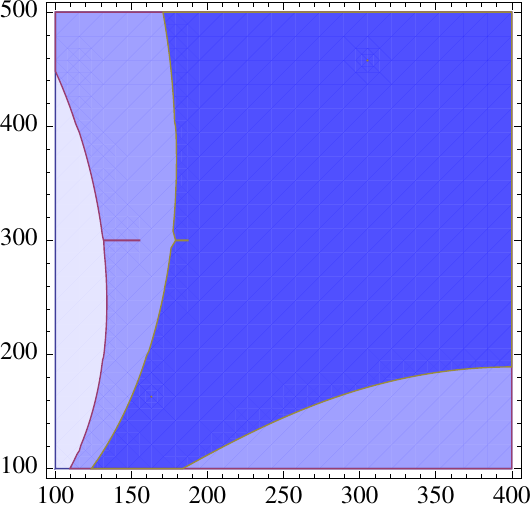}
          }	
\subfigure[$\mu_d = 400$ GeV]{
          \includegraphics[width=0.45\linewidth]{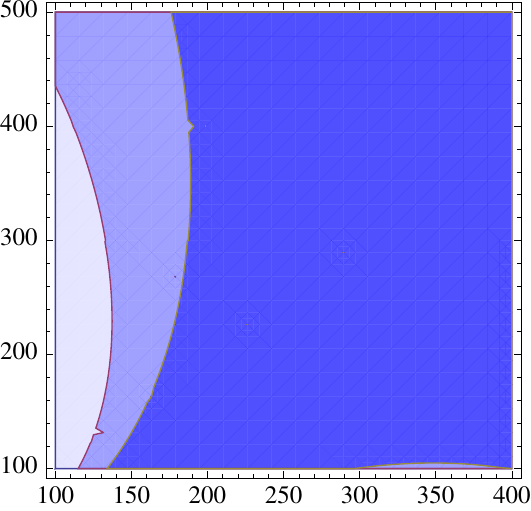}
          } \\ 
\hspace*{0.24\textwidth} $m_{\tilde{l}_1}$~~[GeV]
\hspace*{0.35\textwidth} $m_{\tilde{l}_1}$~~[GeV]
\caption{Same for Fig.~\ref{mu2eR} but with left-handed 
slepton mixing instead.}
\label{mu2eL}
\end{figure}

For both right-handed and left-handed slepton mixing, 
the $Z$-penguin is sub-dominant.  Moreover, for larger values of 
$\tan \beta$, the $Z$-penguin will be even more suppressed, since 
it is directly proportional to $\cos^2 \beta = 1/(1 + \tan^2\beta)$ 
to lowest order, in the limit $M_Z \ll M_N$. 
We show the exclusion plots for $\mu \ra e$ conversion in 
Figs.~\ref{mu2eR} and \ref{mu2eL}.

\subsection{$\mu \ra 3e$}
\label{mu3e-subsec}

Finally, we investigate the decay $\mu^- \ra e^- e^+ e^-$.
The diagrams that contribute to this decay are similar to the process 
$\mu \ra e$ in a nucleus.  While the amplitudes for this decay 
are not enhanced by nuclear factors as in the case of 
$\mu \ra e$ conversion, there is a log enhancement proportional to 
$\log m_\mu/m_e$, arising from an infrared divergence cutoff 
by the electron mass.

All of the diagrams in $\mu \ra 3e$ can be obtained from the 
$\mu \ra e$ conversion diagrams by replacing the quark line by an 
electron line with outgoing $e^+$ and $e^-$. 
All diagrams except the box are the same and will not be discussed here. 
For the box, conservation of $R$-charges enforces both sleptons 
in the loop be of the same ``chirality''. 
The box amplitude for $\mu \ra 3e$ for right-handed sleptons is
\begin{equation}
B_{box}^{R} = \frac{(g' Y_l^R)^4}{16 \pi^2 e^2 } \sin 2\theta_{\tilde{l}} \sum_{i,k=1}^2 \frac{(-1)^{i+1} }{m^2_{\tilde{l}_i}} U_k  j_4\bigg(\frac{M_1^2}{m_{\tilde{l}_i}^2},\frac{M_1^2}{m_{\tilde{l}_i}^2}, \frac{m^2_{\tilde{l}_k}}{m^2_{\tilde{l}_i}}\bigg),
\end{equation}
where $U_1 = \cos^2 \theta_{\tilde{l}}$ and $U_2 = \sin^2 \theta_{\tilde{l}}$. The factor $(-1)^{i+1}$ comes from the super-GIM mechanism. The rate for the decay $\mu \ra 3e$ is 
\begin{eqnarray}
\Gamma_{\mu \ra 3e} & = & \frac{\alpha^2 m_{\mu}^5}{32 \pi} [ (A_{\gamma}^R)^2 - 4 A_{\gamma}^RA^{L}_{\gamma \textrm{dip}} + (A^{L}_{\gamma \textrm{dip}})^2 \bigg(\frac{16}{3}\log \frac{m_{\mu}}{m_e} -\frac{22}{3}\bigg) \nonumber \\
& &{} + \frac{1}{6} (B_{\textrm{box}}^R)^2 +\frac{2}{3}  A_{\gamma}^R B_{\textrm{box}}^R - \frac{4}{3}A^{L}_{\gamma \textrm{dip}}B_{\textrm{box}}^R + \frac{2}{3} F_{RR}^2 +\frac{1}{3} F^2_{RL} \nonumber \\
& &{} +\frac{2}{3}B_{\textrm{box}}^RF_{RR} + \frac{4}{3} A_{\gamma}^R F_{RR} + \frac{2}{3} A_{\gamma}^R F_{RL} - \frac{8}{3}A^{L}_{\gamma \textrm{dip}} F_{RR} - \frac{4}{3} A^{L}_{\gamma \textrm{dip}} F_{RL}],
\end{eqnarray}
where $F_{R\alpha} = A_{Z}^{R} Z_{\alpha}^l$, with $\alpha =$~L,R\@. 
The quantity $Z_{\alpha}$ is part of the electron-$Z$ coupling; 
$Z_L = -1/2 + \sin^2\theta_w$, and $Z_R = \sin^2\theta_w$. 
The branching ratio of this process is obtained by dividing the rate 
by the muon decay rate. Note that the term proportional to 
$(A^{L}_{\gamma \textrm{dip}})^2$ is enhanced by the log term, 
which is divergent in the limit of massless electrons. 
Our result for this divergent term agrees with \cite{DeGouvea:2001mz}.

\begin{figure}
\shortstack[c]{$M_1$ \\ $\mbox{[GeV]}$}
\subfigure[$\mu_d = 100$ GeV]{
          \includegraphics[width=0.45\linewidth]{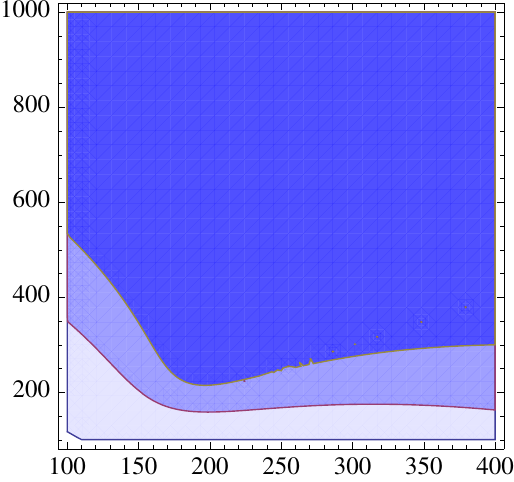}
          }
\subfigure[$\mu_d = 200$ GeV]{
          \includegraphics[width=0.45\linewidth]{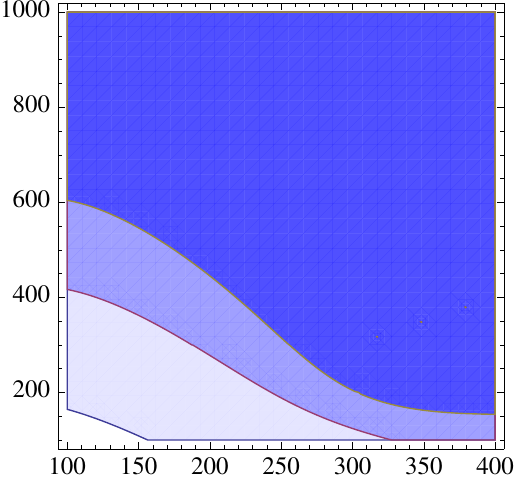}
          } \\
\shortstack[c]{$M_1$ \\ $\mbox{[GeV]}$}
\subfigure[$\mu_d = 300$ GeV]{
          \includegraphics[width=0.45\linewidth]{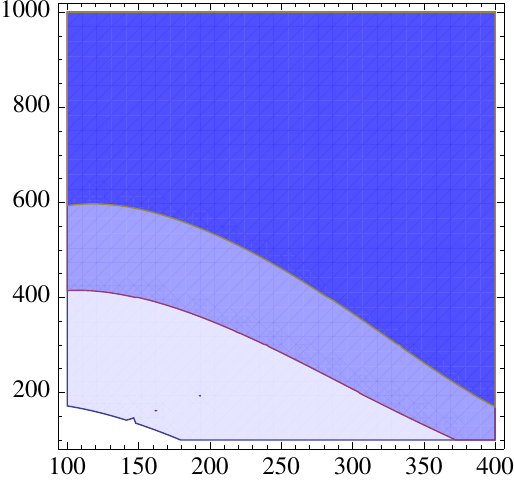}
          }
\subfigure[$\mu_d = 400$ GeV]{
          \includegraphics[width=0.45\linewidth]{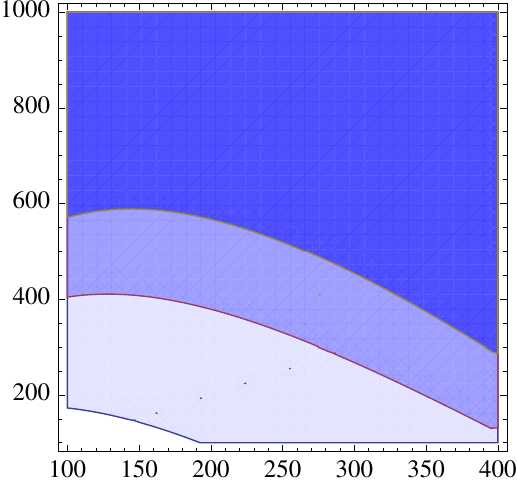}
          } \\ 
\hspace*{0.24\textwidth} $m_{\tilde{l}_1}$~~[GeV]
\hspace*{0.35\textwidth} $m_{\tilde{l}_1}$~~[GeV]
\caption{Regions of the parameter space that satisfy the 
$\mu \ra 3e$ bound at different mixing angles of right-handed sleptons. 
The values of $\sin 2\theta_{\tilde{l}}$ are, from light to dark, 
$0.1$, $0.5$, $1$.}
\label{mu2eeeR}
\end{figure}

\begin{figure}
\shortstack[c]{$M_1$ \\ $\mbox{[GeV]}$}
\subfigure[$\mu_d = 100$ GeV]{
          \includegraphics[width=0.45\linewidth]{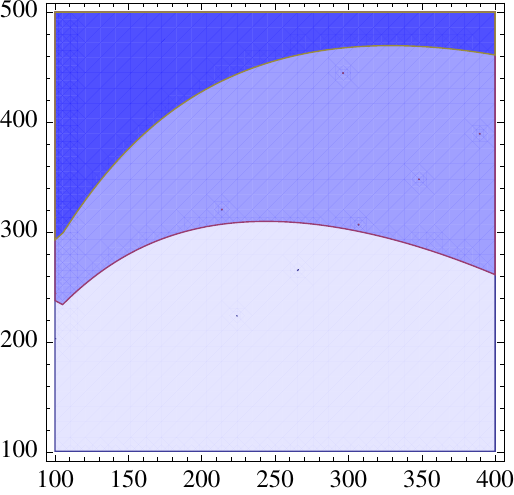}
          }
\subfigure[$\mu_d = 200$ GeV]{
          \includegraphics[width=0.45\linewidth]{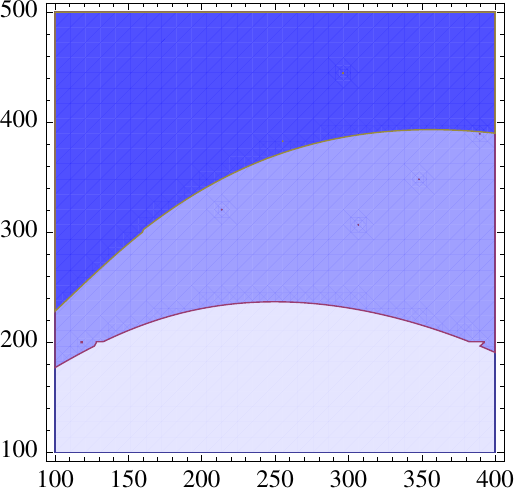}
          } \\	
\shortstack[c]{$M_1$ \\ $\mbox{[GeV]}$}
\subfigure[$\mu_d = 300$ GeV]{
          \includegraphics[width=0.45\linewidth]{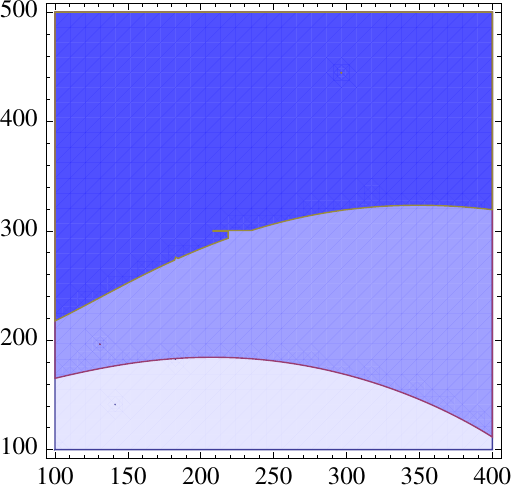}
          }
\subfigure[$\mu_d = 400$ GeV]{
          \includegraphics[width=0.45\linewidth]{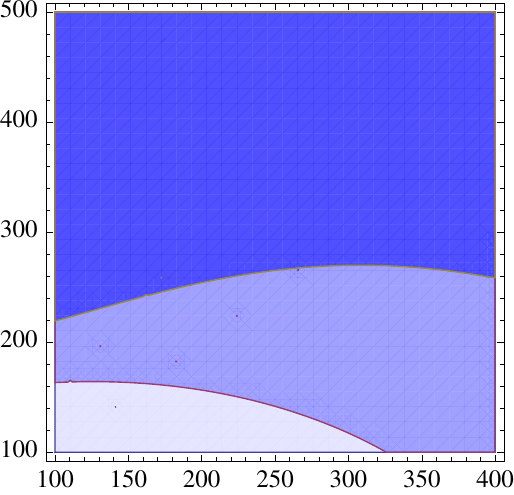}
          } \\ 
\hspace*{0.24\textwidth} $m_{\tilde{l}_1}$~~[GeV]
\hspace*{0.35\textwidth} $m_{\tilde{l}_1}$~~[GeV]
\caption{Same as Fig.~\ref{mu2eeeR} but with left-handed 
slepton mixing.}
\label{mu2eeeL}
\end{figure}

In Figs.~\ref{mu2eeeR},\ref{mu2eeeL} we show the bounds on the 
MRSSM parameter space arising from satisfying the existing 
experimental bound $BR(\mu \ra 3e) < 1.0 \times 10^{-12}$ 
from SINDRUM \cite{Bellgardt:1987du}.
The bounds on the MRSSM parameter space from $\mu \ra 3 e$ are weaker
than the combined bounds from $\mu \ra e\gamma$ and $\mu \ra e$ conversion.

\section{Implications for Flavor Violation Signals at LHC}
\label{lhc-sec}

One of the most interesting implications of the MRSSM is that 
flavor mixing could be at or near maximal throughout virtually
the entire slepton and squark sector \cite{Kribs:2007ac} 
(save only perhaps 
for $\tilde{d}$-$\tilde{s}$ mixing \cite{Blechman:2008gu}).  
For sleptons, this opens up the possibility of observing large 
$\mu$-$e$ mixing at colliders.  Slepton mixing at colliders has been
extensively studied 
\cite{Bityukov:1997ck,Agashe:1999bm,Hisano:2002iy,Kalinowski:2002hv,Deppisch:2004pc,Goto:2004cpa,Hamaguchi:2004ne,Andreev:2006sd,Grossman:2007bd,Feng:2007ke,Nomura:2007ap,Kitano:2008en,Allanach:2008ib,Hirsch:2008dy,Kaneko:2008re,Hisano:2008ng,Feng:2009bs,Feng:2009bd,Kumar:2009sf,Buras:2009sg},
though analyses have generally been relegated to MSSM 
scenarios where the splitting between the $e$,$\mu$ eigenstates 
is very small, so as to satisfy the stringent LFV constraints.
One of the most sensitive techniques to search for $\mu$-$e$ mixing 
is through the decay of a heavier neutralino to a lighter one
through an on-shell slepton.  This decay can arise at a large rate
at the LHC starting with squark and/or gluino production, where
the squark decays to the heavier neutralino and so on, such as
\begin{eqnarray}
\tilde{q} \ra q N_i \; ; \; 
N_i \ra e^\pm/\mu^\pm \tilde{l}^\mp \; ; \; 
\tilde{l}^\mp \ra \mu^\mp/e^\mp N_j \; .
\end{eqnarray}
The distinctive kinematic features in this cascade of 2-body decays 
can be utilized to extract the mass of the slepton through a 
kinematic edge (e.g. \cite{Hinchliffe:1996iu,Abdullin:1998pm,Bachacou:1999zb,Hinchliffe:1999zc,Allanach:2000kt,Aad:2009wy}). 

In light of the bounds on the MRSSM parameter space that we have found
from LFV processes, it is interesting to consider whether large mixing
could still be seen at the LHC\@.  
A detailed collider study is beyond the scope of this paper,
nevertheless we can use our results to uncover characteristic regions
of parameter space where $\sin2\theta_l \sim 1$ simultaneous with
several-hundred GeV sparticles, and thus, where large $\mu \leftrightarrow e$ 
mixing remains within reach of the LHC.   

Closely examining 
Figs.~\ref{mu2eleft}(d),\ref{mu2eL}(d),\ref{mu2eeeL}(d), 
we discover one (small) region in the MRSSM parameter space where 
the left-handed slepton mixing angle can be maximal, $\sin2\theta_l = 1$.  
For this region, and given first and second generation squark
masses to be $1$~TeV (consistent with what was assumed for the 
$\mu \ra e$ conversion numerical results), 
we compute the leading order production cross sections
and decay rates.  We take the wino mass and the right-handed
slepton masses to be 2 TeV for simplicity.  The other gaugino masses 
in this region are $M_1 = 500$~GeV, $\mu_d = 400$~GeV, $\mu_u = 100$~GeV\@. 
The mass spectrum is shown in Table~\ref{spectrum-table}.

\begin{table}[t]
\begin{center}
\renewcommand{\arraystretch}{1.1}
\begin{tabular}{|c|c|c|c|c|c|c|c|c|c|}
\hline
Particle &$\tilde{q}_{L,R}$& $\tilde{g}$&$\neut_3 \simeq \tilde{B}$ &$\charg_2\simeq
\hino_d$&$\neut_2 \simeq \hino_d$&$\tilde{l}_{L2}$ &$\tilde{l}_{L1}$
&$\charg_1\simeq \hino_u$ &$\neut_1 \simeq \hino_u$ \\
\hline
Mass (GeV)  & 1000 & 1000 & 502   & 400   & 400   & 270 & 180 & 100  & 100 \\ 
\hline
\end{tabular}
\end{center}
\caption{Mass spectrum}
\label{spectrum-table}
\end{table}

Using MADGRAPH \cite{Alwall:2007st}, we calculated the leading
order squark and gluino production cross sections at 
LHC with $\sqrt{s} = 14$~TeV center of mass energy for several 
values of the Dirac gluino mass for those production modes allowed 
by $R$-symmetry in Table~\ref{production-sigma-table}.
One important observation made in Ref.~\cite{Kumar:2009sf}
is that, for gluinos less than about 2 TeV, associated
gluino-squark production gives the largest production rate
of squarks.  

\begin{table}[t]
\begin{center}
\begin{tabular}{|c|c|c|c|c|c|}
\hline
$M_{\tilde{g}}$ (TeV) &$\tilde{g}$-$\tilde{q}_{L,R}$ & $\tilde{q}_R$-$\tilde{q}_L$ & $\tilde{q}$-$\tilde{q}^*$ & $\tilde{g}$-$\tilde{g}$& $\sigma (fb)$\\
\hline
1 & 810   & 120 & 50 & 330     & 1300 \\
2 &  36   &  31 & 27 &   1.0   & 95 \\
3 &   2.6 &  11 & 22 &   0.007 & 35 \\
\hline
\end{tabular}
\end{center}
\caption{Leading order production cross sections for squarks and
gluinos at the LHC with $\sqrt{s} = 14$~TeV in the MRSSM.}
\label{production-sigma-table}
\end{table}

The decay rates of the squarks, neutralinos, and charginos,  
computed using BRIDGE \cite{Meade:2007js}, 
can also be computed as a function of the mixing angle $\theta_l$,
shown in Table~\ref{branching-ratios-table}.
\begin{table}[t]
\begin{center}
\begin{tabular}{|c|c|c|}
\hline
Decaying particle &  Decay modes & Branching ratios\\
\hline
$\tilde{q}$ & $q \neut_3$    & $0.99$ \\
\hline
$\neut_3$ & $Z \neut_2$      & $8 \times 10^{-4}$ \\
 & $Z \neut_1$               & $0.12$ \\
 & $\charg_2^- W^+$          & $0.02$ \\
 & $\charg_1^+ W^-$          & $0.22$ \\
 & $ \nu \tilde{\nu}_1$      & $0.19$ \\
 & $ \nu \tilde{\nu}_2$      & $0.13$ \\
 & $ e^- \tilde{l}_{L1}^+$   & $0.19 \cos^2\theta_l$ \\
 & $ \mu^- \tilde{l}_{L1}^+$ & $0.19 \sin^2\theta_l$ \\
 & $ e^- \tilde{l}_{L2}^+$   & $0.13 \sin^2\theta_l$ \\
 & $ \mu^- \tilde{l}_{L2}^+$ & $0.13 \cos^2\theta_l$ \\
\hline
$\tilde{l}_{L1}^+$ & $\charg^+_1 \bar{\nu}$ & $0.11$ \\
 & $\neut_1 e^+$                            & $0.88 \cos^2 \theta_l$ \\
 & $\neut_1 \mu^+$                          & $0.88 \sin^2 \theta_l$ \\
\hline
$\tilde{l}_{L2}^+$ & $\charg^+_1 \bar{\nu}$ & $0.16$ \\
 & $\neut_1 e^+$                            & $0.84 \sin^2\theta_l$ \\
 & $\neut_1 \mu^+$                          & $0.84 \cos^2\theta_l$ \\ 
\hline
\end{tabular}
\end{center}
\caption{Decay branching ratios of the particles involved in the 
cascade decay $\neut_3 \to l^-\tilde{l}_L^+ \to l^- l'^+ \neut_1$
given the MRSSM parameters given in Table~\ref{spectrum-table}.}
\label{branching-ratios-table}
\end{table}%
For the particular point we considered, the first two generations
of squarks decay overwhelmingly into the bino-like neutralino, $\neut_3$. 
The subsequent cascade decays into opposite flavor leptons have
the rates 
$BR(\neut_3 \to e \mu \neut_1) = 0.14 \sin^2 2\theta_l$,
$BR(\neut_3 \to (ee/ \mu \mu) \neut_1) = 0.27 (\sin^4\theta_l+\sin^4\theta_l)$.
If the gluino mass is 1 TeV, for example, then the $\tilde{g}\tilde{q}$ 
production leads to a total cross section of about 1 pb.  With maximal 
slepton mixing, the cross section for opposite sign $e \mu$ events 
is expected to be of order 100 fb.  Extracting this signal from
background, particularly given the potentially problematic 
technique of flavor-subtraction, remains challenging. 
(See Ref.~\cite{Kumar:2009sf} for a discussion of signal 
plus background analysis of a non-minimal $R$-symmetric model.)

Just as in the MSSM, one can search for the kinematic endpoint in the 
invariant mass distribution of the leptons.  In the MRSSM, however,
the two slepton mass eigenstates are not near one another,
and so two distinct and well-separated kinematic edges 
could in principle be extracted.  This would be a striking signal
of slepton flavor violation in the MRSSM.  Note also that 
the electric charges of the leptons in this decay are fixed by 
the conversation of $R$-charges.  For example, the anti-neutralino
$\neut_3^c$ can decay into $l^+\tilde{l}_L^-$, the decay into the same 
final state for $\neut_3$ is forbidden.

\section{Discussion}
\label{discussion-sec}

\begin{figure}
\shortstack[c]{$M_1$ \\ $\mbox{[GeV]}$}
        \subfigure[$\mu_d = 100$ GeV]{
                \includegraphics[width=0.45\linewidth]{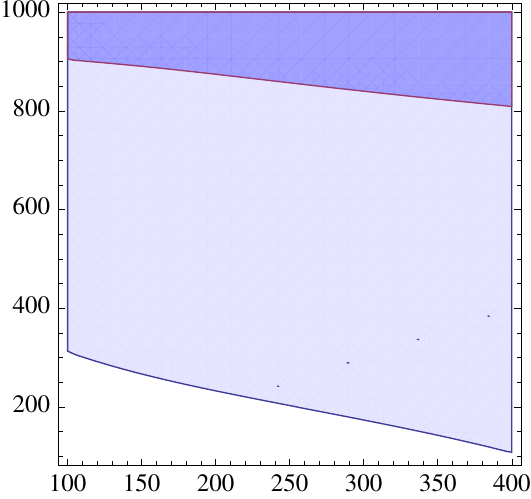}
                }
        \subfigure[$\mu_d = 200$ GeV]{
                \includegraphics[width=0.45\linewidth]{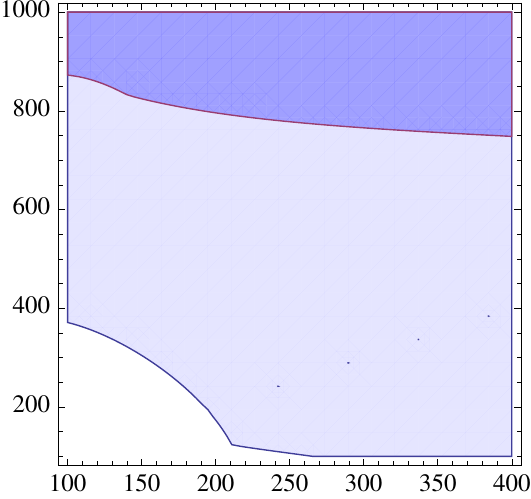}
                } \\
\shortstack[c]{$M_1$ \\ $\mbox{[GeV]}$}
        \subfigure[$\mu_d = 300$ GeV]{
                \includegraphics[width=0.45\linewidth]{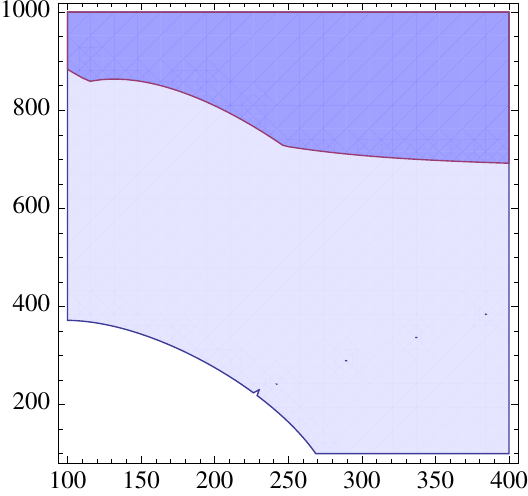}
                }
        \subfigure[$\mu_d = 400$ GeV]{
                \includegraphics[width=0.45\linewidth]{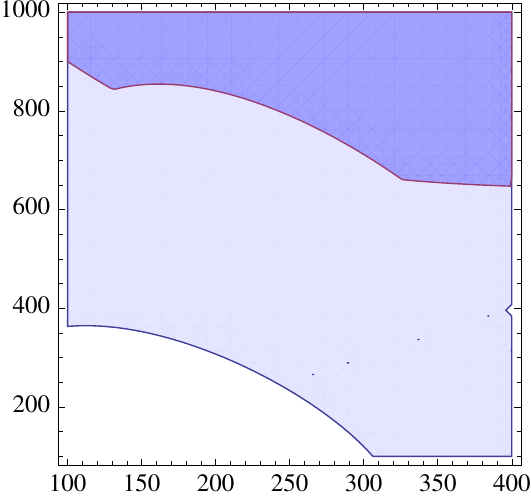}
          } \\ 
\hspace*{0.24\textwidth} $m_{\tilde{l}_1}$~~[GeV]
\hspace*{0.35\textwidth} $m_{\tilde{l}_1}$~~[GeV]
\caption{Regions allowed in the parameter space by combining the three
constraints for right handed sleptons. The constraint from $\mu \ra 3e$ 
is always less severe than the other two processes in the 
parameter space shown.}
\label{regionRH}
\end{figure}

\begin{figure}
\shortstack[c]{$M_1$ \\ $\mbox{[GeV]}$}
\subfigure[$\mu_d = 100$ GeV]{
          \includegraphics[width=0.45\linewidth]{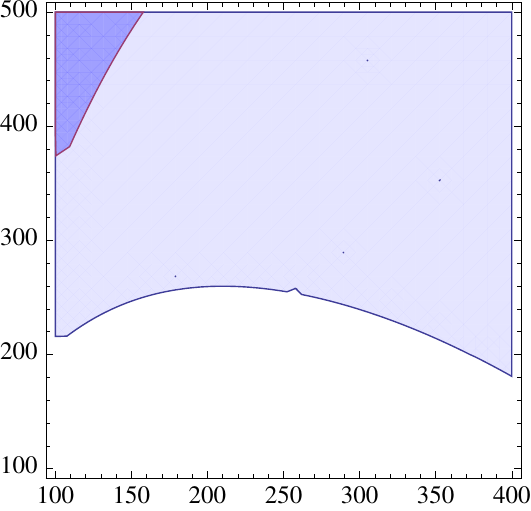}
          }
\subfigure[$\mu_d = 200$ GeV]{
          \includegraphics[width=0.45\linewidth]{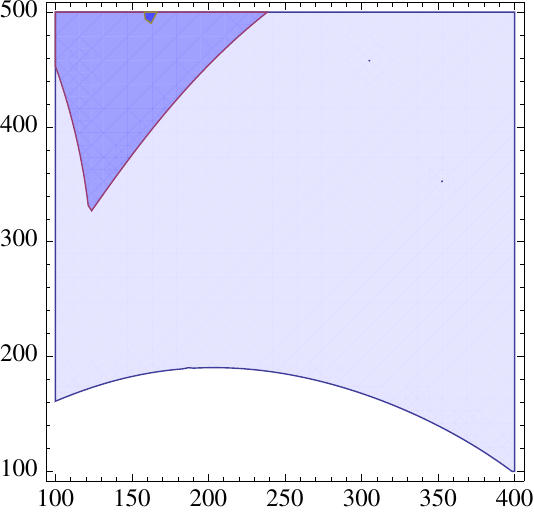}
          } \\
\shortstack[c]{$M_1$ \\ $\mbox{[GeV]}$}
\subfigure[$\mu_d = 300$ GeV]{
          \includegraphics[width=0.45\linewidth]{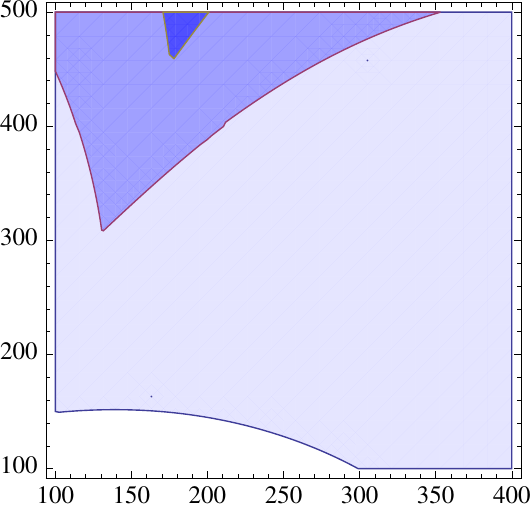}
          }
\subfigure[$\mu_d = 400$ GeV]{
          \includegraphics[width=0.45\linewidth]{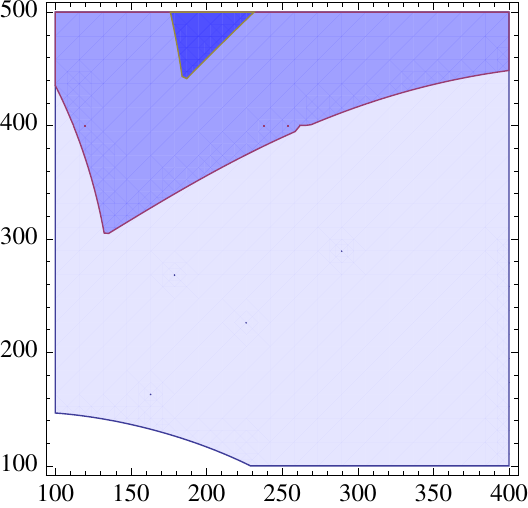}
          } \\ 
\hspace*{0.24\textwidth} $m_{\tilde{l}_1}$~~[GeV]
\hspace*{0.35\textwidth} $m_{\tilde{l}_1}$~~[GeV]
\caption{Same as Fig.~\ref{regionRH} but for left handed sleptons.  
Similar to the right handed case, the constraint from $\mu \ra 3e$ 
is also less severe than the other two processes in the parameter space 
shown.}
\label{regionLH}
\end{figure}

We have calculated the constraints on $\mu \leftrightarrow e$ mixing
in the MRSSM from the flavor violating processes 
$\mu \ra e \gamma$, $\mu \ra e$ conversion, and $\mu \ra 3e$.
Given the simplifications stated in Sec.~\ref{simp-sec}, 
we explored LFV in the MRSSM as a function of the parameters
$M_1, \mu_d, m_{\tilde{l}}$, and $\sin 2\theta_{\tilde{l}}$ 
within the sub-TeV range.  Given the heavier slepton mass set to be 
$m_{\tilde{l}_2} = 1.5 m_{\tilde{l}_2}$, 
we found that the bound from $\mu \ra 3e$ is always less severe 
than the bounds derived from either $\mu \ra e \gamma$ or 
$\mu \ra e$ conversion.  We show the overlapping regions
allowed by all constraints in Figs.~\ref{regionRH},\ref{regionLH}.

For right-handed slepton mixing, $\mu \ra e$ conversion in 
gold nuclei provides the most severe constraint -- it completely 
rules out maximal mixing (compare Fig.~\ref{regionRH} with 
Fig.~\ref{mu2eright}). 
The situation is qualitatively different for 
left-handed mixing -- the most severe bound in this case comes 
from $\mu \ra e \gamma$, as dominant amplitudes 
(charge radius and magnetic dipole) of $\mu \ra e$ conversion interfere 
destructively and opens up a large region in parameter space 
that satisfies the experimental bounds.  From Fig.~\ref{mu2eleft} 
for $\mu \ra e \gamma$, one sees that maximal mixing is allowed 
in regions where the bino mass is $\sim 500$~GeV at $\mu_d = 200$~GeV, 
with a moderate splitting between sleptons.  The results suggest
that the most likely observation of large slepton flavor violation
signals at the LHC will occur in the left-handed sector.

Finally, is interesting to consider how the bounds on slepton
flavor mixing angles will change as the constraints on LFV 
are strengthened.  This is most easily understood by recognizing
that all of our bounds are proportional to $\sin^2 2 \theta_{\tilde{l}}$. 
In other words, the boundary of the allowed regions are contours 
of constant $BR_{\textrm{bound}} / \sin^2 2 \theta_{\tilde{l}}$, 
where $BR_{\textrm{bound}}$ is the bound on the branching ratio 
of a process.  In plotting the allowed regions of parameter space
in the previous sections of the paper, we used of course the 
current experimental bound.  Suppose that in some future 
experiment the bounds are improved, say by a factor of $100$.  
Then, the boundary of the region that satisfy this new bound 
for $\sin^2 2 \theta_{\tilde{l}} = 0.1$ is the same 
as the boundary for the current bound with 
$\sin^2 2 \theta_{\tilde{l}} = 1$.  

\section*{Acknowledgments}

We thank A.~Martin and T.~Roy for discussions as well as invaluable
assistance in running MADGRAPH with the MRSSM model. 
We also thank D.~Tucker-Smith 
for clarifications of some of the results in Ref.~\cite{Kumar:2009sf}.
This work was supported in part by the Department of Energy 
under contract DE-FG02-96ER40969.


\begin{appendix}
\refstepcounter{section}

\section*{Appendix~\thesection:~~Gaugino and Slepton Structure}
\label{app-sec}

To discuss the neutralino masses and interactions more 
quantitatively, we define the $\psi_{B}$ and $\psi_{\hino_d}$ 
to the be fermion $R$-partners of $\bino$ and $\hino_d^0$, 
respectively. Then we form the Dirac bino and Higgsino spinors 
and their charge conjugates,
\begin{equation}
\neut_{\tilde{B}} = \binom{\psi_B}{\tilde{B}^{\dagger}}, 
\quad \neut_{\hino_d} = \binom{\hino_d^0}{\psi_{\hino_d}^{\dagger}}, 
\quad \neut^c_{\tilde{B}} = \binom{\tilde{B}}{\psi_B^{\dagger}}, \quad 
\neut^c_{\hino_d} = \binom{\psi_{\hino_d}}{\hino_d^{0\dagger}}.
\end{equation}
We can also see that the Dirac spinor $\neut$ has an $R$-charge $-1$, 
whereas $\neut^c$ has an $R$-charge $+1$.  The gaugino mass matrix, 
$M_{\neut}$, is shown in the mass term below
\begin{equation}
(\bar{\neut}_B, \bar{\neut}_{\hino_d}) \left( 
		\begin{array}{cc}
		M_1 & -\cos\beta \sin\theta_W M_Z \\
		0 & \mu_d \\
		\end{array}
		\right)  \binom{P_L \neut_B}{P_L \neut_{\hino_d}} + 
                h.c.
\label{mixingmatrix}
\end{equation}

The mass matrix is diagonalized by a bi-orthogonal transformation; 
the diagonalized neutralino mass matrix, 
$M_{\neut}^D = O_L^T M_{\neut} O_R$, obey 
$(M_{\neut}^D)^2 = O_L M_{\neut} (M_{\neut})^T (O_L)^T 
= O_R (M_{\neut})^T M_{\neut} (O_R)^T$, 
where $O_{(L,R)}$ are the orthogonal matrices that diagonalize 
the mass matrix. In this definition, the $\bino$ and $\hino^0_d$ 
content of the $i$-th neutralino $\neut_i$ are, $O_{Li1}$ and $O_{Ri2}$, 
respectively.

We consider mixing between selectrons and smuons only, 
parameterized as follows:
\begin{equation}
\left(
\begin{array}{c}
\tilde{l}_{1} \\
\tilde{l}_{2}
\end{array}
\right)_{L,R} =
\left(
\begin{array}{cc}
  \cos\theta_{\tilde{l}}& \sin\theta_{\tilde{l}} \\
 -\sin\theta_{\tilde{l}} &  \cos\theta_{\tilde{l}}
\end{array}
\right)_{L,R}
\left(
\begin{array}{c}
\tilde{e} \\
\tilde{\mu}
\end{array}
\right)_{L,R},
\label{sleptonmixing}
\end{equation}
where $\tilde{l}_i$ represents the sleptons in the mass-eigenstate basis.

Then slepton flavor violation comes from the interaction terms 
between a sfermion, $\tilde{f}_i$, a neutralino, $\neut_i$, and a fermion $f_i$:
\begin{equation}
-\tilde{f}_{L\alpha}^* \bar{\neut}_i(U_{L\alpha\beta}^{\dagger} [O_{Li1} G_L f_{L\beta} +O_{Ri2} y_f f_{R\beta}])  -\tilde{f}_{R\alpha}^* \bar{\neut}^c_i(U_{R\alpha\beta}^{\dagger}[O_{Li1} G_R f_{R\beta} +O_{Ri2} y_f f_{L\beta}]) + h.c.
\end{equation}
where $U_{L,R}$ are the slepton mixing matrices in Eq.~(\ref{sleptonmixing}). 
The coupling constants are
\begin{eqnarray}
G_{L,R} &=& \sqrt{2} g' Y_{f_{(L,R)}}, \quad \textrm{and} \\
y_f     &=& \frac{g' m_f}{\sqrt{2} M_Z \sin\theta_w \cos \beta}.
\end{eqnarray}

The subscript $i$ on the (s)fermion denotes its generation, 
subscripts $L$ and $R$ denote the chirality, with $\alpha$ and $\beta$ 
being the flavor indices. The hypercharge of a fermion $f$ 
is denoted by $Y_f$. From the above interaction terms we see that 
$\tilde{f}_R$ and $\tilde{f}_L$ have different $R$-charges; $-1$ and $+1$, 
respectively.

The $Z$-boson only couples to Higgsinos. The $Z \neut \neut$ 
interaction term is
\begin{equation}
\frac{g}{2\cos\theta_w}Z_{\mu}[\bar{\neut}_i \gamma^{\mu}(O_{Ri2}O_{Rj2} P_L + O_{Li2}O_{Lj2}P_R) \neut_j].
\end{equation}
One can also write the $Z\neut\neut$ coupling in terms of $\neut^c$,
\begin{equation}
-\frac{g}{2\cos\theta_w}Z_{\mu}[\bar{\neut^c}_i \gamma^{\mu}(O_{Li2}O_{Lj2} P_L + O_{Ri2}O_{Rj2}P_R) \neut^c_j].
\end{equation}

Examining the neutralino mixing matrix in Eq.~(\ref{mixingmatrix}), 
the lightest gaugino receives a negative shift, $-\Delta < 0$, 
and so the lightest neutralino has mass 
$M_{\neut_1} = \mu_d - \Delta < m_{\charg_1}$, and thus 
the lightest gaugino is a neutralino.

\end{appendix}


\end{document}